\documentclass[twocolumn,showpacs,nofootinbib,preprintnumbers,amsmath,amssymb,superscriptaddress]{revtex4}
\usepackage{epsfig}
\pdfoutput=1

\usepackage[english]{babel}
\usepackage{amsmath,amssymb,amsthm,amsfonts}

\usepackage{ulem}
\usepackage{ mathrsfs }
\usepackage{slashed}
\usepackage{graphicx}
\usepackage{subfigure}
\usepackage{color,rotating}
\usepackage{dsfont}
\usepackage{setspace}
\usepackage{verbatim}
\usepackage{fancyhdr}
\definecolor{refcol}{RGB}{0,0,205}
\usepackage{microtype}
\usepackage[colorlinks,linkcolor=refcol,citecolor=refcol,urlcolor=refcol]{hyperref}
\usepackage{breakurl}

\usepackage{tabularx}

\newcommand{\beq}{\begin{equation}}
\newcommand{\eeq}{\end{equation}}
\newcommand{\beqa}{\begin{eqnarray}}
\newcommand{\eeqa}{\end{eqnarray}}
\newcommand{\bfc}{\begin{figure}[h!]\begin{center}}
\newcommand{\efc}{\end{center}\end{figure}}
\newcommand{\nn}{\nonumber}

\def\Fig#1{Fig.~\ref{#1}}

\graphicspath{{./figures/}}

\newcommand{\bea}{\begin{eqnarray}}
\newcommand{\eea}{\end{eqnarray}}
\newcommand{\Eq}[1]{Eq.~(\ref{#1})}

\newcommand*{\mn}{{\mu\nu}}

\def\eq#1{(\ref{#1})}
\def\Eq#1{Eq.~(\ref{#1})}
\newcommand {\apgt} {\ {\raise-.5ex\hbox{$\buildrel>\over\sim$}}\ }
\newcommand {\aplt} {\ {\raise-.5ex\hbox{$\buildrel<\over\sim$}}\ }

\def\s0#1#2{\mbox{\small{$ \frac{#1}{#2} $}}}
\def\0#1#2{\frac{#1}{#2}}

\def\CN{{\mathcal N}}

\newcommand{\be}{\begin{eqnarray}}
\newcommand{\ee}{\end{eqnarray}}

\newcommand{\Nf}{N_{f}}
\newcommand{\Nc}{N_{\rm{c}}}

\definecolor{blue}{rgb}{0,0,1}

\definecolor{green}{rgb}{0,1,0}

\definecolor{red}{rgb}{1,0,0}

\definecolor{gray}{rgb}{.5,.5,.5}

\definecolor{darkgreen}{rgb}{.0,.5,.0}

\begin{document}
\title{From Quarks and Gluons to Hadrons: Chiral Symmetry Breaking in Dynamical QCD}

\author{Jens Braun}
\affiliation{Institut f\"ur Kernphysik
  (Theoriezentrum), Technische Universit\"at Darmstadt, Schlo\ss gartenstra\ss e 2, D-64289
  Darmstadt, Germany} \affiliation{ExtreMe Matter Institute EMMI, GSI,
  Planckstra{\ss}e 1, D-64291 Darmstadt, Germany} 
\author{Leonard Fister}
\affiliation{Institut de Physique Th\'eorique, CEA Saclay, F-91191 Gif-sur-Yvette, France}

\author{Jan M.~Pawlowski}
\affiliation{Institut f\"ur Theoretische
  Physik, Universit\"at Heidelberg, Philosophenweg 16, 69120
  Heidelberg, Germany} \affiliation{ExtreMe Matter Institute EMMI, GSI,
  Planckstra{\ss}e 1, D-64291 Darmstadt, Germany} 

\author{Fabian Rennecke}
\affiliation{Institut f\"ur Theoretische
  Physik, Universit\"at Heidelberg, Philosophenweg 16, 69120
  Heidelberg, Germany} \affiliation{ExtreMe Matter Institute EMMI, GSI,
  Planckstra{\ss}e 1, D-64291 Darmstadt, Germany}


\begin{abstract} 
  We present an analysis of the dynamics of two-flavour QCD in the
  vacuum. Special attention is paid to the transition from the high
  energy quark-gluon regime to the low energy regime governed by
  hadron dynamics. This is done within an functional renormalisation
  group approach to QCD amended by dynamical hadronisation
  techniques. The latter allow us to describe conveniently the
  transition from the perturbative high-energy regime to the
  nonperturbative low-energy limit without suffering from a
  fine-tuning of model parameters. In the present work, we apply these
  techniques to two-flavour QCD with physical quark masses and show
  how the dynamics of the dominant low-energy degrees of freedom
  emerge from the underlying quark-gluon dynamics.
\end{abstract} 

\maketitle

\section{Introduction}
For an accurate first-principles description of the dynamics of QCD, a
reliable inclusion of hadronic states is of great importance. This
holds in particular for an approach aiming at the hadron spectrum or
the phase structure of QCD at finite density. In the present work on
two-flavour QCD we develop a theoretical framework for taking into
account the fluctuation dynamics of quarks, gluon and hadrons. This
approach is based on previous functional renormalisation group
studies~\cite{Braun:2008pi,Braun:2009gm} and a related quantitative
study in the quenched limit \cite{Mitter:2014wpa}. The present work
and \cite{Mitter:2014wpa} are first works within a collaboration
(fQCD) aiming at a quantitative functional renormalisation group
framework for QCD \cite{FRG-QCD}. This framework allows to dynamically
include hadronic states as they emerge from the microscopic quark and
gluon degrees of freedom.

We use the functional renormalisation group (FRG) approach for QCD,
for reviews see
\cite{Litim:1998nf,Berges:2000ew,Pawlowski:2005xe,Schaefer:2006sr,%
  Gies:2006wv,Pawlowski:2010ht,%
  Rosten:2010vm,Braun:2011pp,vonSmekal:2012vx,JMP_QM}, and
\cite{Alkofer:2000wg,Roberts:2000aa,Fischer:2006ub,
  Fischer:2008uz,Binosi:2009qm,Maas:2011se,Boucaud:2011ug} for reviews
on related work. In order to describe the transition from quarks and
gluons to hadrons, we extend the dynamical hadronisation technique (or
rebosonisation), introduced in
Refs.~\cite{Gies:2001nw,Gies:2002hq,Pawlowski:2005xe,Floerchinger:2009uf}. For
the first time, this technique is applied here to dynamical
two-flavour QCD with physical quark masses. It is shown how the
dominant hadronic low-energy degrees of freedom and their dynamics
emerge from the underlying quark-gluon dynamics.  The hadronisation
technique, as further developed in the present work, already applied
in Ref.~\cite{Mitter:2014wpa} in a quantitative study of quenched
QCD. In the latter work, a large number of interaction channels has
been taken into account, aiming at full quantitative precision.  Here,
we exploit the results from \cite{Mitter:2014wpa} as well as results
on the scale-dependent glue sector of Yang-Mills theory from
Refs.~\cite{Fischer:2008uz,Fister:2011uw,FP}. This enables us to
concentrate on the RG flows of the most relevant couplings from a more
phenomenological point of view, paying special attention to
unquenching effects.

In summary, the aim of this work is threefold: Firstly, we aim at a
detailed understanding of the fluctuation physics in the transition
region between the high energy quark-gluon regime to the low energy
hadronic regime. Secondly, we want to initiate the quest for the minimal
set of composite operators that have to be taken into account for
reaching (semi-)quantitative precision, while keeping the study
analytic. This deepens the understanding of the
fluctuation physics by only taking into account the relevant
operators. Moreover, it is also of great interest for low energy effective
models. Thirdly, we discuss full unquenching effects in terms of the
matter back-coupling to the glue sector that is important for QCD
regimes with dominant quark fluctations such as QCD at high densities
or many flavours. 

The paper is organised as follows: In Sect.~\ref{sec:action} we
introduce the ansatz for the quantum effective action which we are
considering in the present work. The general framework of dynamical
hadronisation is then discussed in detail in Sect.~\ref{sec:qfl},
where we also give a discussion of the RG flow in the gauge sector and
the role of the quark-gluon vertex. Our results for two-flavour QCD
are then presented in Sect.~\ref{sec:results}. While our analysis
suggests that the use of dynamical hadronisation techniques only
yields mild quantitative corrections in low-energy model studies, its
use is indispensable from both a qualitative and a quantitative point
of view for a unified description of the dynamics of QCD on all
scales. Our conclusions are given in Sect.~\ref{sec:conc}. Some
technical details as well as a brief discussion about the effect of
dynamical hadronisation on low-energy models are discussed in the
appendices.

\section{The effective action}\label{sec:action}

Our aim is to describe two-flavour QCD in $d=4$ Euclidean dimensions
at vanishing temperature and density in a vertex expansion. The
starting point is the microscopic gauge fixed QCD action. Thus, we
include the quark-gluon, three- and four-gluon vertices as well as the
ghost-gluon vertex and the corresponding momentum-dependent
propagators. Four-quark interactions are dynamically generated at
lower scales and we therefore take the scalar-pseudoscalar channel
into account in our truncation. This is by far the dominant four-quark
channel, as it exhibits quark condensation, see
\cite{Mitter:2014wpa}.

On even lower energy scales, bound state
degrees of freedom appear and eventually become dynamical. To properly
take this into account, we introduce a scale-dependent effective
potential $V_k$ which includes arbitrary orders of mesonic
self-interactions. Since dynamics in this sector is dominated by the
lightest mesons, we restrict our analysis to pions and the sigma-meson
and their corresponding momentum-dependent propagators. We therefore
assume a strong axial anomaly, i.e. $U(1)_A$ is maximally broken. As a
consequence, the meson sector in the chiral limit exhibits an $O(4)$
flavor symmetry. Note that this is also reflected in the four-quark
interaction: the scalar-pseudoscalar channel $\sim \lambda_{q,k}$ is
invariant under $SU(2)_V \!\times\! SU(2)_A$ but violates $U(1)_A$
symmetry, see \eq{eq:trunc}. Explicit chiral symmetry breaking is
included via a source term $-c \sigma$. It is directly related to a
finite current quark mass and, as a consequence, non-zero pion
masses. This implies that we have a chiral crossover transition rather
than a second order phase transition. The meson sector is coupled to
the quark sector by a field-dependent Yukawa coupling
$h_k(\phi^2)$. That way, arbitrarily high orders of quark-antiquark
multi-meson correlators are included \cite{Pawlowski:2014zaa}. We
elaborate on the physics picture in Sect.~\ref{sec:results}.

The key
mechanism to consistently describe the dynamical generation of bound
state degrees of freedom in this work is dynamical hadronisation, and
is discussed in Sect.~\ref{sec:dynhad}.  In summary, this yields the
following scale-dependent effective action,
\begin{align}\nonumber
  \Gamma_k =& \int_x\biggl\{ \frac{1}{4}F^a_{\mu \nu} F^a_{\mu \nu} +
 Z_{c,k} \bar c^a \partial_\mu D^{ab}_\mu c^b + \frac{1}{2\xi} (\partial_\mu
  A_\mu^a)^2\\\nonumber &
  +{Z}_{q,k}\, \bar{q} \left(\gamma_\mu D_\mu
  \right) q-\lambda_{q,k} \left[(\bar{q}\, T^0 q)^2-(\bar{q} \gamma_5
    \vec{T} q)^2\right]\biggr.\\\nonumber &+\biggl. h_{k}(\phi^2)
  \left[ \bar{q} (i\gamma_5\vec{T}\vec{\pi}
    +T^0\sigma) q \right]+\frac{1}{2} Z_{\phi,k} (\partial_{\mu} \phi)^2\\
  &+V_k(\rho)-c\sigma \biggr\}+\Delta\Gamma_{\text{glue}} \,,
 \label{eq:trunc}
\end{align}
with the $O(4)$ meson field $\phi\!=\! (\sigma,\vec{\pi})$ and
$\rho\!=\!  \phi^2/2$. $D_\mu\!=\!\partial_\mu-i Z^{1/2}_{A,k}\, g_{k}
A_\mu^a t^a$ is the Dirac operator, with the strong coupling
$g_k=\sqrt{4 \pi \alpha_{s,k}}$ and the gluonic wave-function
renormalisation $Z_{A,k}$. With this definition the covariant
derivative $D_{\mu}$ is renormalisation group invariant. The last term
in the first line, $\Delta\Gamma_{\text{glue}}$, stands for the
non-trivial ghost-gluon, three-gluon and four-gluon vertex
corrections, for further details see Sect.~\ref{sec:gauge} and in
particular Eq.~\eq{eq:deltagammaglue}. The full momentum
  dependence of the pure gauge sector is taken into account in the
  gluon and ghost dressing functions $Z_{A,k}$ and $Z_{c,k}$. This is
  crucial for the correct IR behaviour of the gauge sector.

Due to asymptotic freedom the effective action at the initial cutoff
scale $\Lambda$ relates to the classical (gauge-fixed) QCD action,
\begin{align}\label{eq:truncLambda}
  \nonumber\Gamma_{k\to\Lambda} \simeq &\int_x\biggl\{
  \frac{1}{4}F^a_{\mu \nu} F^a_{\mu \nu} + \bar{q} \left(\gamma_\mu
    D_\mu + m_q^{\text{\tiny UV}}
  \right) q \\
  & \qquad\quad + \bar c^a \partial_\mu D^{ab}_\mu c^b +
  \frac{1}{2\xi} (\partial_\mu A_\mu^a)^2 \biggr\}\,.
\end{align} 
The quark mass~$m_q^{\text{\tiny UV}}$ at the UV scale~$\Lambda$ is
directly related to the coupling $c$ in Eq.~\eqref{eq:trunc}. The other couplings appearing in our
ansatz~\eqref{eq:trunc} for the effective action are generated
dynamically in the RG flow.

In this work, we use Hermitian gamma
matrices so that
\begin{equation}
\{\gamma_{\mu},\gamma_{\nu}\}=2 \delta_{\mu \nu} \mathds{1}\,.
\end{equation}
The commutator for the $SU(N_c)$ generators reads $[t^a,t^b] \!=\! i f^{abc}
t^c$ and, hence, the trace is positive, $\text{Tr}\, t^a
t^b \!=\! \frac{1}{2} \delta^{ab}$. $\vec{T}$ are the $SU(N_f)$ generators
and $T^0\! =\!\frac{1}{\sqrt{2 N_f}} \mathds{1}_{N_f\times N_f}$.  For
the field strength tensor we use the relation
\begin{align}\label{eq:Fren}
F_{\mu\nu} & =  \0{i}{Z^{1/2}_{A,k} g_{k}  } [ D_{\mu}\,,\,D_{\nu}]\\ \nonumber
& = 
Z^{1/2}_{A,k}\, t^a \left( \partial_{\mu} A^a_{\nu}-\partial_{\nu} A^a_{\mu}+
Z^{1/2}_{A,k} \,g_{k}   f^{a
b c} A^b_{\mu} A^c_{\nu}\right)\,.
\end{align}
For more details on the gauge part of our truncation see
Sect.~\ref{sec:gauge}.
All masses, wave-function renormalisations and couplings are scale-dependent. The
scalar potential and the Yukawa coupling are expanded about a
scale-independent point $\kappa$, $\partial_t\kappa=0$. As shown in
\cite{Pawlowski:2014zaa} this yields a rapid convergence of the
expansion
\begin{align}\nonumber 
V_k(\rho) &= \sum_{n=1}^{N_V} \frac{v_{n,k}}{n!}\left( \rho-
\kappa \right)^n,\\[2ex]
h_k(\rho) &= \sum_{n=0}^{N_h} \frac{h_{n,k}}{n!}\left( \rho-\kappa \right)^n\,.
\label{potential}\end{align}
Note that the quark and meson mass
functions (two-point functions at vanishing momentum) depend on the
meson fields. The masses are given by the mass functions evaluated at 
the physical minimum $\rho_{0,k} = \sigma_0^2/2$ of $V_k(\rho)-c\sigma$, 
\begin{align}\nonumber 
m_{q,k}^2 &= \frac{1}{2} h_k^2(\rho_{0,k}) \rho_{0,k}\,,\\[2ex]\nonumber 
m_{\pi,k}^2 &= V'(\rho_{0,k})\,,\\[2ex]
m_{\sigma,k}^2 &= V'(\rho_{0,k})+2\rho_{0,k}V''(\rho_{0,k})\,,
\label{eq:masses}
\end{align}
where $m_{q,k}$ is the constituent quark mass. The current quark mass
$m_q^{\text{\tiny UV}}$ is related to the symmetry breaking source $c$ via the mass
function at the ultraviolet scale,
\begin{align}\label{eq_currm}
m_q^\text{\tiny UV} = \frac{h_\Lambda}{2 v_{1,\Lambda}} c\,, 
\end{align} 
while $c$ does not occur explicitly in the flow equation as it is the
coefficient of a one-point function. This entails that the flows of
the effective action in the chiral limit and that in QCD with
non-vanishing current quark masses agree, see also
\cite{Pawlowski:2014zaa}. The difference solely relates to the
solution of the equation of motion for the $\sigma$-field,
\begin{align}\label{EoMsigma}
\left. \0{\delta\Gamma_{k=0}}{\delta \sigma}\right|_{\sigma=\sigma_{\rm EoM}}=0\,.
\end{align}
If expanding the flow in powers of the mesonic fields as done in the
present work, the expansion point has to be close to $\sigma_{\rm
  EoM}$, such that it is within the radius of convergence of the
expansion. 

\section{Quantum fluctuations}\label{sec:qfl}
Quantum fluctuations are computed with the functional renormalisation
group. For QCD related reviews and corresponding low-energy models, we
refer the reader to
Refs.~\cite{Litim:1998nf,Berges:2000ew,Pawlowski:2005xe,Schaefer:2006sr,Gies:2006wv,Pawlowski:2010ht,%
  Rosten:2010vm,Braun:2011pp,vonSmekal:2012vx,JMP_QM}. A consistent
description of the dynamical transition from quark-gluon degrees of
freedom to hadronic degrees of freedom is achieved by the dynamical
hadronisation technique. Loosely speaking, it is a way of storing
four-quark interaction channels, which are resonant at the chiral
phase transition, in mesonic degrees of freedom and therefore allows
for a unified description of the different degrees of freedom
governing the dynamics at different momentum scales.

\subsection{Functional RG \& dynamical hadronisation}\label{sec:dynhad} 
The starting point of the functional renormalisation group is the
scale-dependent effective action $\Gamma_\Lambda$ at a UV-cutoff scale
$\Lambda$. In the case of QCD, $\Lambda$ is a large,
perturbative energy scale and correspondingly $\Gamma_\Lambda$ is the
microscopic QCD action with the strong coupling constant and the
current quark masses as the only free parameters of the theory. From there, quantum
fluctuations are successively included by integrating out momentum
shells down to the RG scale $k$. This yields the scale-dependent
effective action $\Gamma_k$, which includes all fluctuations from momentum
modes with momenta larger than $k$. By lowering $k$ we resolve the
macroscopic properties of the system and eventually arrive at the full
quantum effective action $\Gamma = \Gamma_{k=0}$. The RG evolution of
the scale-dependent effective action is given by the Wetterich
equation \cite{Wetterich:1992yh}, which in the case of QCD with
$\Phi=(A,q,\bar{q},c,\bar c,\phi)$ reads 
\begin{align}\label{fleq}
  \nonumber&\partial_t\Gamma_k[\Phi]=\\ \nonumber
 &\quad\frac{1}{2}\text{Tr}\bigl(G_{AA,k}[\Phi]\cdot \partial_t
  R_k^A\bigr)-\text{Tr}\bigl(G_{c\bar c,k}[\Phi]\cdot \partial_t
  R_k^c\bigr)\\
  & \quad-\text{Tr}\bigl(G_{q\bar{q},k}[\Phi]\cdot \partial_t
  R_k^q\bigr)+\frac{1}{2}\text{Tr}\bigl(G_{\phi\phi,k}[\Phi]\cdot \partial_t
  R_k^\phi\bigr).
\end{align}
Here, the regulator functions $R_k^{\Phi_i}(p)$ can be understood as
momentum-dependent masses that introduce the suppression of infrared
modes of the respective field $\Phi_i$, and are detailed in
App.~\ref{app:thres}. The derivative $\partial_t$ is the total
derivative with respect to the RG scale $t=\ln( k/\Lambda)$ with some
reference scale $\Lambda$. The traces sum over discrete and continuous
indices of the fields, including momenta and species of fields. The
first line on the right hand side of \eq{fleq} is the flow in the pure
glue sector, the second line creates the matter
fluctuations. $G_k[\Phi]$ denote the scale and field-dependent full
propagators of the respective fields, e.g. for the quarks
\begin{align}\label{eq:prop}
 G_{q\bar{q},k}[\Phi]=\left(\frac{\delta^2\Gamma_k[\Phi]}{\delta
      q(-p)\delta \bar{q}(p)}+R_k^q\right)^{-1}.
\end{align}
In the following, we will not encounter mixed two-point
functions. Hence, it is sufficient to define these expression for the
combinations quark--anti-quark, meson-meson, gluon-gluon (both
transverse) and ghost--anti-ghost. For the rest of the manuscript, we
drop the redundant second field-index for the two-point functions and
the propagators. In a slight abuse of notation we define the scalar
parts of the two-point functions of the quark, meson, gluon and ghost
as
\begin{align}\label{eq:Gam2}
\begin{split}
  \Gamma^{(2)}_{q,k}(p)\equiv \frac{\delta^2 \Gamma_k[\Phi]}{\delta
    q(-p)\delta\bar{q}(p)}\,, \ \Gamma^{(2)}_{\phi,k}(p)\equiv
  \frac{\delta^2 \Gamma_k[\Phi]}{\delta
    \phi(-p)\delta\phi(p)}\,,\\
  \Gamma^{(2)}_{A,k}(p)\equiv \frac{\delta^2 \Gamma_k[\Phi]}{\delta
    A(-p)\delta A(p)}\,, \ \Gamma^{(2)}_{c,k}(p)\equiv \frac{\delta^2
    \Gamma_k[\Phi]}{\delta c(-p)\delta\bar{c}(p)}\,.
\end{split}
\end{align}
With this we define the corresponding wave-function renormalisations
and (scalar parts of the) propagators
\begin{align}\label{eq:G}
\begin{split}
  Z_{\Phi_i,k}(p) &= \Delta \Gamma^{(2)}_{\Phi_i,k}(p)/
  \Delta S^{(2)}_{\Phi_i}(p)\Bigr|_\text{scalar part}\,,\\
  G_{\Phi_i,k}(p) &= \left(Z_{\Phi_i,k}(p) \Delta S^{(2)}_{\Phi_i}+
    R_k^{\Phi_i}(p)\right)^{-1}\Bigr|_\text{scalar part}\,,
\end{split}
\end{align}
with $\Phi_i = q,\phi,A$ or $c$. The scalar part is the coefficient of
the tensor structure of the expressions above. In \eq{eq:G} we have
$\Delta \Gamma^{(2)}_{\Phi_i,k}(p)=
\Gamma^{(2)}_{\Phi_i,k}(p)-\Gamma^{(2)}_{\Phi_i,k}(0)$ for all fields
except for the gluon, where $\Delta \Gamma^{(2)}_{A,k}(p)=
\Gamma^{(2)}_{A,k}(p)$. The same holds true for $\Delta
S^{(2)}_{\Phi_i}$. At $k=0$ and the fields set to their vacuum
expectation value, $G_{\Phi_i,k=0}(p) $ is the full propagator. The
above definitions are exemplified with the full gluon
propagator,
\begin{align}\label{eq:gluepropdef}
G_{A,k}^{ab}(p) = \frac{1}{Z_{A,k}(p)\, p^2 + R_k^A}\, \Pi^\perp \delta^{ab}\,,
\end{align}
with the transversal projection operator $\Pi^\perp$, see
\eq{eq:Pi}. For our calculations, we use four-dimensional Litim
regulators $R_k$, \cite{Litim:2000ci}, for details see
App.~\ref{app:thres}.

In the infrared regime of QCD, the dynamical degrees of freedom are
hadrons, while quarks and gluons are confined inside hadrons. This
entails that a formulation in terms of local composite fields with
hadronic quantum numbers is more efficient in this regime. Note that these 
composite fields are directly related to hadronic observables at their poles. 

Let us illustrate this at the relevant example of the
scalar-pseudoscalar mesonic multiplet at a given cutoff scale $k$. At
a fixed large cutoff scale, where the mesonic potential $V_k(\rho)$ is
assumed to be Gau\ss ian, we can resort to the conventional
Hubbard-Stratonovich bosonisation: the local part of the
scalar--pseudo-scalar channel of the four-quark interaction with
coupling $\lambda_{q,k}$, see the second line in \eq{eq:trunc}, can be
rewritten as a quark-meson term, see the third line in \eq{eq:trunc},
on the equations of motion for $\phi$, that is $\phi_{\rm EoM}$. This
leads to
\begin{align}\label{eq:hs}
\lambda_{q,k}=\frac{h_{k}^2}{2 v_{1,k}}\,,  \qquad \qquad \phi_{j,\rm EoM}
= \frac{h_{k} }{ v_{1,k} } \bar{q} \boldsymbol{\tau}_j q\,, 
\end{align}
where $v_1$ is the curvature mass of the mesonic field and
$\boldsymbol{\tau}=(\gamma_5 \vec{T},iT^0)$, $j\!\in\!
\{1,2,3,4\}$. Note that \eq{eq:hs} is only valid for $Z_\phi\equiv 0$
and a Gau\ss ian potential $V_k(\rho) = v_1\rho$. Moreover,
mis-counting of degrees of freedom may occur from an inconsistent
distribution of the original four-fermi interaction strength to the
Yukawa coupling and the four-fermi coupling. The dynamical
hadronisation technique used in the present work, and explained below,
resolves these potential problems.

One advantage of the bosonised formulation concerns the direct access
to spontaneous chiral symmetry breaking via the order parameter
potential $V_k(\rho)$: spontaneous symmetry breaking is signaled by
$v_1=0$ at the symmetry breaking scale $k_{\chi}$ which relates to a 
resonant four-quark interaction. It also facilitates the access to the 
symmetry-broken infrared regime. 

\begin{figure}[t]
\begin{center}
\includegraphics[width=0.75\columnwidth]{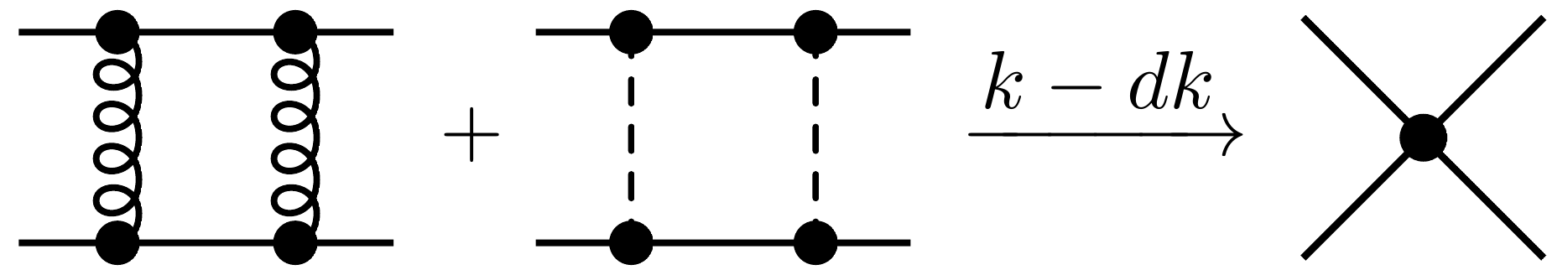}
  \caption{Re-generation of four-quark interactions from the RG-flow.}\label{fig:box}
\end{center}
\end{figure}

Let us now assume that we have performed the above complete
bosonisation at some momentum scale $k\gg k_\chi$. There, the above
conditions for the bosonisation in \eq{eq:hs} are valid. Hence, we can
remove the four-fermi term completely in favour of the mesonic Yukawa
sector. However, four-quark interactions are dynamically re-generated
from the RG flow via quark-gluon and quark-meson interactions, see
\Fig{fig:box}. 

Indeed, these dynamically generated contributions dominate due to the
increase of the strong coupling $\alpha_{s,k}$ for a large momentum
regime, leading to a quasi-fixed point running of the Yukawa coupling,
see Refs.~\cite{Gies:2001nw,Gies:2002hq,Mitter:2014wpa} and Sec.~\ref{sec:results}. Thus, even though $\lambda_{q,k}$ was exactly
replaced by $m_{\phi,k}$ and $h_k$ at a scale $k\gg k_\chi$, there is
still a non-vanishing RG-flow of $\lambda_{q,k}$ at lower
scales. Note, however, that we have explicitly checked that this is
only a minor quantitative effect as long as one considers low-energy
effective models, see App.~\ref{app:lowhad}.

In summary, it is not
possible to capture the full dynamics of the system in the quark-gluon
regime with the conventional Hubbard-Stratonovich bosonisation. As a
consequence, with the conventional bosonisation, the scale where
composite fields take over the dynamics from fundamental quarks and
gluons is not an emergent scale generated by the dynamics of QCD, but
is fixed by hand by the scale where the Hubbard-Stratonovich
transformation is performed.

In the present approach we employ dynamical hadronisation instead of
the conventional bosonisation. It is a formal tool that allows for a
unified description of dynamically changing degrees of freedom and
consequently is not plagued by the shortcomings of conventional
bosonisation discussed above. It has been introduced in
\cite{Gies:2001nw} and was further developed in
\cite{Gies:2002hq,Pawlowski:2005xe,Floerchinger:2009uf}. The
construction works for general potentials $V_k(\rho)$ (more precisely 
general $\Gamma_k[\Phi]$), and implements the idea of bosonising multi-fermion
interactions at every scale $k$ rather just at the initial
scale. Consequently, the resulting fields of this bosonisation
procedure, i.e. the mesons, become scale-dependent and can be viewed
as hybrid fields: while they act as conventional mesons at low
energies, they encode pure quark dynamics at large energy scales.

Here we follow the dynamical hadronisation set-up put forward in
\cite{Pawlowski:2005xe} and outline the derivation of the flow
equation in the presence of scale-dependent meson fields. The starting
point is the functional integral representation of the scale-dependent
effective action $\Gamma_k$ with scale-dependent meson fields. To this
end, we define the dynamical superfield
$\hat\Phi_k=(\hat\varphi,\hat\phi_k)$, where the microscopic fields
are combined in $\hat\varphi=(\hat A_\mu,\hat q,\hat{\bar{q}},\hat
c,\hat{\bar c})$ and the scale-dependent meson fields, in our case
pions and the sigma meson, are represented by the $O(4)$ field
$\hat\phi_k=(\hat{\vec{\pi}}_k,\hat{\sigma}_k)$. The path integral
representation of $\Gamma_k$ reads 
\begin{align}\label{eq:paga}
 e^{-\Gamma_k[\Phi_k]}=&\int\! \mathcal{D} \hat{\varphi} \,
  \exp\!\left\{-S[\hat{\varphi}]-\Delta S_k[\hat{\Phi}_k]\right.\\ \nonumber
&+\left.\frac{\delta (\Gamma_k
      +\Delta S_k)}{\delta \Phi_k}
    (\hat{\Phi}_k-\Phi_k)+\Delta S_k[\Phi_k]\right\},
\end{align}
where we defined the expectation value of the fields
$\Phi_k=\langle\hat\Phi_k\rangle$ and used 
\begin{align}
J=\frac{\delta \left ( \Gamma_k+\Delta S_k\right)}{\delta \Phi_k} \quad
\text{and} \quad \Delta S_k[\Phi_k]=\frac{1}{2} \Phi_k {R}_k \Phi_k\,.
\end{align}
Note that the functional integral in \eq{eq:paga} contains only the
fundamental fields $\hat\varphi$ of QCD. Composite operators such as
the (scale dependent) mesons are introduced via corresponding source
terms in the Schwinger functional, see
\cite{Pawlowski:2005xe}.  

To arrive at the evolution equation for $\Gamma_k[\Phi_k]$, we take
the scale derivative $\partial_t=k\frac{d}{dk}$ of
Eq.~\eq{eq:paga}. The RG evolution of the scale-dependent composite
meson fields is of the form
\begin{align}\label{eq:dtphi}
\partial_t \hat\phi_k = \dot A_k \bar{q} \boldsymbol{\tau}q+\dot B_k \hat\phi_k\,.
\end{align}
The first part of this equations reflects the bound state nature of
the mesons. The second part corresponds to a general rescaling of the
fields. The coefficients $\dot A_k$ and $\dot B_k$, which we call
hadronisation functions, are specified below. Note that the right hand
side of (\ref{eq:dtphi}) only involves the quark mean fields
$q\!=\!\langle\hat q\rangle$, $\bar q\!=\!\langle\hat{
  \bar{q\,}}\rangle$. An explicit solution to this equation is given
by
\begin{align}\label{eq:phisol}
\hat\phi_k = C_k\, e^{B_k}\bar q \boldsymbol{\tau}q\,,
\end{align}
with $\dot A_k = \dot C_k e^{B_k}$. This reflects the quark-antiquark
nature of the meson. Eq.~\eq{eq:dtphi} leads to the following
identity for the flow of the hadronisation field
\begin{align}\label{eq:dtphi2}
  \langle\partial_t\hat\phi_k\rangle=\dot A_k \bar{q}
  \boldsymbol{\tau}q+\dot B_k \phi_k\,,
\end{align}
and furthermore $\langle\partial_t\hat\phi_k\rangle \!=\! \partial_t
\phi_k$. Taking \eq{eq:dtphi} into account, the scale derivative of
\eq{eq:paga} gives a modified version of the flow equation
\eq{fleq}. While the gauge and quark parts of the equation remain
unchanged, the mesonic part now reads:
\begin{align}\label{eq:hadflow}
\begin{split}
  \bigl.\partial_t\bigr|_\phi\Gamma_k[\Phi_k]=&\frac{1}{2}\text{Tr}
  \left[ G_{\phi\phi,k}[\Phi]\cdot\left(\partial_t R_k^\phi+2R_k^\phi
      \dot{B}_k\right) \right]\\[2ex]
  &-\text{Tr}\left[
    \frac{\delta\Gamma_k[\Phi]}{\delta\phi_i}\left(\dot A_k \bar{q}
      \boldsymbol{\tau_i}q+\dot B_k \phi_i\right) \right].
\end{split}
\end{align}
The first line of (\ref{eq:hadflow}) corresponds to the mesonic part
of the flow equation (\ref{fleq}) with a shift in the scale derivative
of the regulator owing to the part of $\partial_t\phi_k$ which is
proportional to $\phi_k$ itself. Note that (\ref{eq:hadflow}) remains
valid for the more general flow of the super-field
\cite{Pawlowski:2005xe}
\begin{align} 
\partial_t\hat\Phi_{i,k}=\dot A_{ij,k}\!\cdot\!
F_{j,k}[\Phi_k]+\dot B_{ij,k}[\Phi_k]\hat\Phi_{j,k}\,,
\end{align}
where $F[\Phi_k]$ is any functional of the mean super-field
$\Phi_k$. We emphasise that the one-loop nature of the flow equation
is not spoiled as long as $\partial_t\hat\Phi_{i,k}$ is at most linear
in the quantum field $\hat\Phi_{i,k}$. It can, in fact, be an
arbitrary function of the mean fields $\Phi_{i,k}$ without altering
the properties of the flow equation. The meson regulator has the form
(see App.~\ref{app:thres})
\begin{align}
R_k^\phi(p^2)=Z_{\phi,k} p^2 r_B(p^2/k^2)\,,
\end{align}
and its corresponding scale derivative can conveniently be
written as
\begin{align}
  \partial_t R_k^\phi(p^2)=\left( \bigl.\partial_t\bigr|_{
      Z}-\eta_{\phi,k} \right) R_k^\phi(p^2)\,,
\end{align}
with the anomalous dimension of the scale-dependent mesons, 
\begin{align}
\eta_{\phi,k}=-\frac{\partial_t Z_{\phi,k}}{ Z_{\phi,k}}\,.
\end{align}
This choice of the regulator functions implies that the flow equations
of RG-invariant quantities only contain the anomalous dimension which
stems from the scale derivative of the regulator whereas the
wave-function renormalisations drop out completely. With this, we can
rewrite (\ref{eq:hadflow}) into:
\begin{align}
  \nonumber \bigl.\partial_t\bigr|_\phi\Gamma_k[\Phi_k]=
  &\frac{1}{2}\text{Tr}\left[ G_{\phi\phi,k}[\Phi]\cdot
    \left(\bigl.\partial_t\bigr|_{Z}-(\eta_{\phi,k}-
      2\dot{B}_k)\right)R_k^\phi \right]\\
  &-\text{Tr}\left[
    \frac{\delta\Gamma_k[\Phi]}{\delta\phi_i}\left(\dot A_k \bar{q}
      \boldsymbol{\tau_i}q+\dot B_k \phi_i\right) \right].
\end{align}
It is now obvious that the first line of the modified flow equation
above gives the original flow equations without scale-dependent
fields, but with a shifted meson anomalous dimension:
\begin{align}\label{eq:etashift}
\eta_{\phi,k}\rightarrow\eta_{\phi,k}-2\dot{B}_k\,.
\end{align}
The other coefficient, $\dot B_k$, in (\ref{eq:dtphi}) is at our
disposal, and we may use it to improve our truncation. 

The second line of (\ref{eq:hadflow}) induces additional contributions
in particular to the flows of the four-quark and the Yukawa coupling,
owing to the particular ansatz we made for $\partial_t\phi_k$. This
allows us to specify the hadronisation procedure: we choose the
coefficient $\dot A_k$ such that the flow of the four-quark
interaction $\lambda_{q,k}$ vanishes within our truncation,
$\partial_t \lambda_{q,k}=0$. This way, all information about the
multi-quark correlations are stored in the flow of the Yukawa
coupling. Thus, $h_k$ encodes the multi-quark correlations in the
quark-gluon regime and the meson--constituent-quark correlations in the
hadronic regime, including a dynamical transition between these
different regimes.

\subsection{Hadronised flow equations}\label{sec:hadflows}

In the following we specify the hadronisation procedure and give the
resulting modified flow equations of the scale-dependent parameters of
the truncation (\ref{eq:trunc}). These modifications are given by
explicitly evaluating the second line of (\ref{eq:hadflow}). Note that
the explicit form of the modified flow equations depends on the details 
of our projection procedures, see also App.~\ref{app:flows}. 

In the following, we rescale all fields with their respective
wave-function renormalisation, $\bar\Phi=\sqrt{Z_{\Phi,k}}\Phi$ and
introduce the RG-invariant parameters
\begin{align}\label{eq:barg}
  &\bar{g}_k=\frac{g_k}{Z_{q,k} Z_{A,k}^{1/2}}\,,\quad
  \bar{\lambda}_{q,k}=
  \frac{\lambda_{q,k}}{Z_{q,k}^2},\quad \bar c_k = \frac{c}{Z_{\phi,k}}\,,\\ \nonumber
  &\bar{\lambda}_{n,k}=\frac{\lambda_{n,k}}{Z_{\phi,k}^n}\,,\quad
  \bar h_{n,k}=\frac{h_{n,k}}{Z_{q,k} Z_{\phi,k}^{(2n+1)/2}}\,,\quad \bar\kappa_k = Z_{\phi,k}\kappa\,.
\end{align}
Note that the parameters defined in \eq{eq:barg} do scale with the
infrared cutoff scale $k$, but are invariant under general
RG-transformations (reparameterisations) of QCD. For example, $\bar
g_k$ is nothing but the running strong coupling.  The RG-invariant
dimensionless masses are defined accordingly as
\begin{align}\label{eq:mbars}
  \bar m_{q,k}=\frac{m_{q,k}}{k\, Z_{q,k}}\quad \text{and} \quad \bar
  m_{\pi/\sigma,k}=\frac{m_{\pi/\sigma,k}}{k\, Z_{\phi,k}^{1/2}}\,.
\end{align}
Note that we rescale mesonic parameters with the wave-function
renormalisation $Z_{\phi,k}$ of the scale-dependent mesons
$\phi_k$. The constant source $c$ as well as the expansion point
$\kappa$ have only canonical running after rescaling, given only by
the running of $ Z_{\phi,k}$, see Eq.~(\ref{eq:kandc}). Consequently,
we also rescale the hadronisation functions and, in addition, define
them to be dimensionless:
\begin{align}
  \dot{\bar A}_k = k^2
  Z_{\phi,k}^{1/2}Z_{q,k}^{-1}\dot{A}_k\,,\qquad \dot{\bar B}_k
  =\dot{B}_k\,.
\end{align}
With this, we proceed now to the modified flow equations of these
RG-invariant quantities.

For the flow of the four-quark interaction $\bar\lambda_{q,k}$ we find:
\begin{align}\label{eq:hadh}
\begin{split}
  \bigl.\partial_t\bigr|_{\phi}\bar\lambda_{q,k} =&\, 2\,\eta_{q,k}
  \bar\lambda_{q,k}+\bigl.\partial_t\bar\lambda_{q,k}
  \bigr|_{\eta_{\phi,k}\rightarrow\tilde\eta_{\phi,k}-2\dot{\bar B}_k}\\
  &+\left( \bar h_k(\bar\rho)+2\bar\rho\bar{h}_k'(\bar\rho)\frac{4
      N_f N_c-1}{2N_f N_c+1} \right)\dot{\bar A}_k\, .
\end{split}
\end{align}
Here, $\partial_t\bar\lambda_{q,k}$ denotes the flow without dynamical
hadronisation which is specified in App.~\ref{app:flows}. As
already discussed above, this contribution is subject to a shift in
the meson anomalous dimension, indicated by
$\eta_{\phi,k}\rightarrow\eta_{\phi,k}-2\dot{\bar B}_k$.

Following the discussion in the previous section, we choose $\dot{\bar
  A}_k$ such that the flow of $\bar\lambda_{q,k}$ vanishes. This is
achieved by the following choice:
\begin{align}\label{eq:A}
\begin{split}
  \dot{\bar A}_k=&-\left( \bar h_k(\bar\rho)+2\bar\rho
    \bar{h}_k'(\bar\rho)\frac{4 N_f N_c-1}{2N_f N_c+1} \right)^{-1}\bigl.\\
  &\times\partial_t\bar\lambda_{q,k}\bigr|_{\eta_{\phi,k}
    \rightarrow\eta_{\phi,k}-2\dot{\bar B}_k}\,.
\end{split}
\end{align}
Together with the initial condition $\bar\lambda_{q,\Lambda}=0$, this yields
\begin{align}
\bigl.\partial_t\bigr|_{\phi}\bar\lambda_{q,k}=0.
\end{align}

The flow of the Yukawa coupling assumes the following form:
\begin{align}\label{eq:hadh1}
 \nonumber \bigl.\partial_t\bigr|_{\phi}\bar h_k =& \left(\!
    \eta_{q,k}+\frac{1}{2} \eta_{\phi,k} \!\right)\bar
  h_k+\bigl.\partial_t\bar{h}_{k}
  \bigr|_{\eta_{\phi,k}\rightarrow\tilde\eta_{\phi,k}-2\dot{\bar B}_k}\\
  &-\frac{1}{k^2}\left( p^2+\bar{V}_k'(\bar\rho) \right)\dot{\bar
    A}_k-\left(\bar h_k +2\bar\rho\bar h_k'\right)\dot{\bar B}_k\, ,
\end{align}
where $\bar h_k=\bar h_k(\bar\rho)$ is implied and
$\partial_t\bar{h}_{k}$ is specified in App.~\ref{app:flows}. 
From Eq.~\eqref{eq:A}, it is now clear that the flow of the quark interaction
and, therefore, all information about the multi-quark correlations
within our truncation is incorporated into the flow of the hadronised
Yukawa coupling. 

It is left to specify the hadronisation function $\dot{\bar B}_k$,
which also enters (\ref{eq:hadh1}). We see from
  Eq.~\eq{eq:phisol} that it corresponds to a phase factor of the
  hadronisation field. It can be used to improve the current
approximation by absorbing a part of the momentum-dependence of the
mesonic wave-function renormalisation and the Yukawa coupling. This
will be discussed elsewhere. Here, we use
\begin{align}\label{eq:dotB=0} 
\dot B_k\equiv 0\,,
\end{align}
for the sake of simplicity. We see that our hadronisation procedure enforces a vanishing
four-quark interaction. The effect of four-quark correlations is then
stored in the Yukawa coupling, which now serves a dual purpose: while
it captures the current-quark self-interactions in the quark-gluon
regime, it describes the meson--constituent-quark in the hadronic
regime.

\subsection{Gauge sector}\label{sec:gauge}
In this section, we discuss the gauge sector of the truncation given
in \eq{eq:trunc}. Most importantly, this permits to distinguish the
quark-gluon coupling from pure gluodynamics. This directly signals the
transition from the perturbative quark-gluon regime at large momenta,
where all couplings scale canonically, to the hadronic regime where
non-perturbative effects are dominant.

The couplings induced from three-point functions play a dominant role
in the description of interactions. Hence, we solve the flow equations
for all three-point functions in QCD, the quark-gluon, three-gluon and
ghost-gluon vertices. In addition, the effects from the four-gluon
vertex are important \cite{Fischer:2008uz,Fister:2011uw,FP}. Thus, we
employ an ansatz which has proven to be accurate in previous studies
\cite{Fister:2011uw, FP}. For the computation presented here, we take
the gluon and ghost propagators from pure gauge theory as input
\cite{Fischer:2008uz,Fister:2011uw,FP} and augment them by unquenching
effects. In the perturbative domain this procedure is accurate, as the
error is order $\alpha_{s,k}^2$. At scales below the confinement
transition the gluon is gapped and therefore decouples from the
dynamics.

Perturbation theory gives a direct relation between the number of
gluon legs $m$ attached to the vertex $\Gamma^{(n)}$ and the order in
the strong coupling, $\Gamma^{(n)}\sim (4 \pi
\alpha_{s,k})^{m/2}$. Nevertheless, the RG running is different,
although purely induced by the external legs attached. Their
wave-function renormalisations cancel exactly those from the
propagators, see \eq{eq:simpleA} below. As a result of this
truncation, the flow equations for couplings depend on the anomalous
dimensions only.

In this analysis we restrict ourselves to classical tensor structures
of the gauge action $S[\Phi]$. Omitting colour and Lorentz indices for
clarity, we parametrise the quark-gluon, three- and four-gluon and the
ghost-gluon vertices as
\begin{align}\label{eq:vertices}
\begin{split}
  \Gamma_k^{(\bar{q} A q)}&=Z_{A,k}^\frac{1}{2} Z_{q,k}\, g_{\bar{q} A q,k}\ S^{(3)}_{\bar{q} A q}\,,\\
  \Gamma_k^{(A^3)}&=Z_{A,k}^\frac{3}{2}\, g_{A^3,k}\ S^{(3)}_{A^3}\,,\\
  \Gamma_k^{(A^4)}&=Z_{A,k}^2\, g^2_{A^4,k}\ S^{(4)}_{A^4}\,,\\
  \Gamma_k^{(\bar{c} A c)}&=Z_{A,k}^\frac{1}{2} Z_{c,k}\, g_{\bar{c} A
    c,k}\ S^{(3)}_{\bar{c} A c}\,.
\end{split}
\end{align}
The classical tensor structures $S^{(n)}_{\Phi_1...\Phi_n}$ are
obtained from \eq{eq:truncLambda} by 
\beq S^{(n)}_{\Phi_1...\Phi_n} = \left. \frac{\delta^{n}
    \Gamma_{\Lambda}}{{\delta\Phi_1 \ldots \delta \Phi_n}}
\right|_{g_k=1}\,, \eeq
where we have omitted indices for clarity.
\begin{figure}[t]
\begin{center}
  \includegraphics[width=.95\columnwidth]{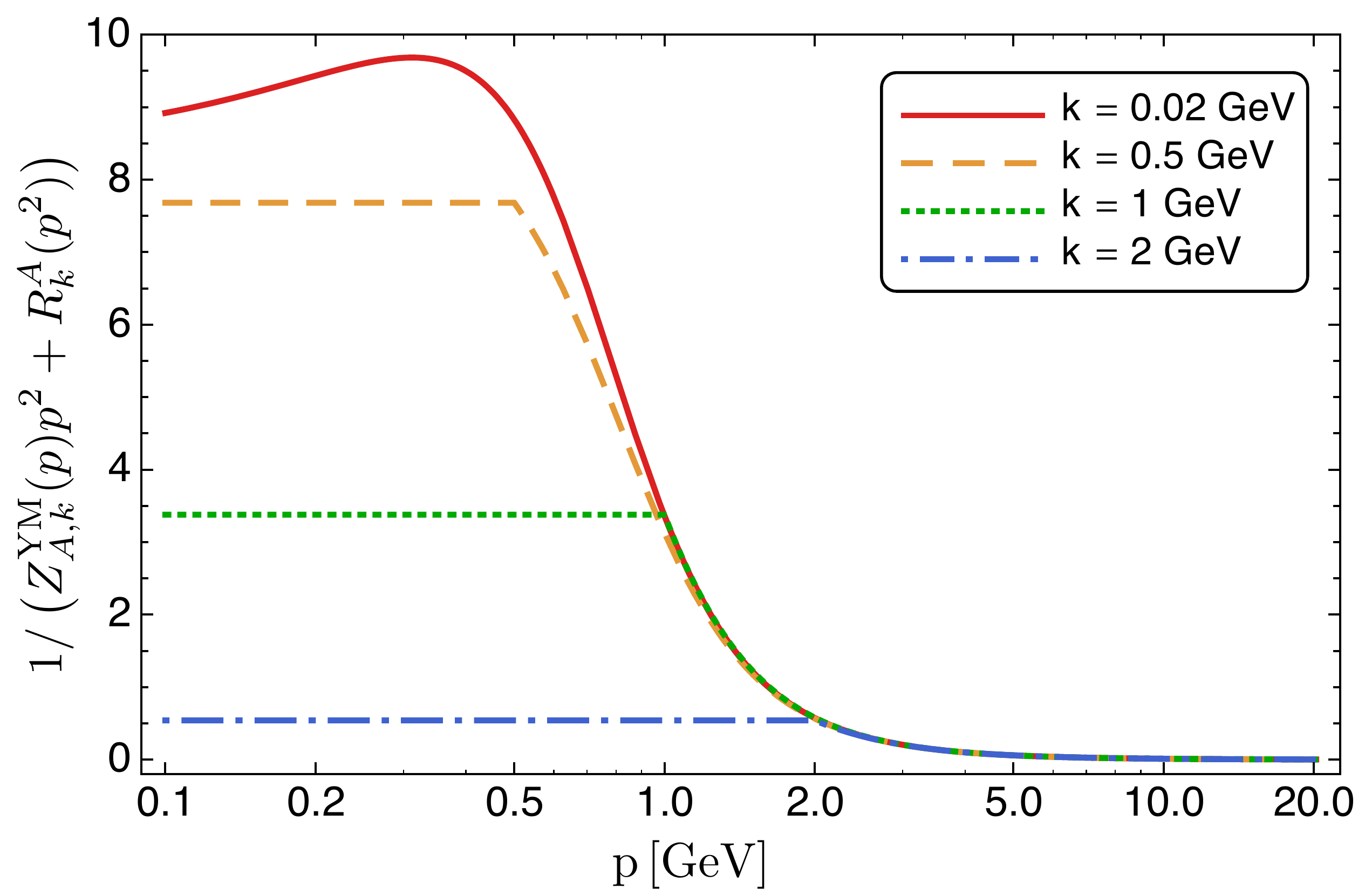}
  \caption{The regulated gluon propagator from pure Yang-Mills theory as a
    function of the momentum for various $k$. We use this as an
    external input for our QCD computations.}\label{fig:gluein}
\end{center}
\end{figure}
In this work, we use as input the gluon/ghost two-point functions
$\Gamma_{A/c,k}^{(2),{\rm YM}}(p)$ computed in
\cite{Fischer:2008uz,Fister:2011uw,FP} for pure Yang-Mills theory, 
\begin{align}
\begin{split}
\Gamma_{A,k}^{(2),\text{YM}} &= Z_{A,k}^{\text{YM}}(p)\, p^2 \Pi^\perp\,,\\
\Gamma_{c,k}^{(2),\text{YM}} &= Z_{c,k}^{\text{YM}}(p)\, p^2\,,
\end{split}
\end{align}
where the identity matrix in adjoint color space is implied. 
In Figs.~\ref{fig:gluein} and \ref{fig:ghin} we show this input as a
function of the momentum $p$ for various $k$. Note that we show
  the regulated gluon propagator and ghost dressing function with optimized regulators $R_{k}^{A/c}
  \!=\!  (Z_{A/c,k}^{\text{YM}}(k)\, k^2-Z_{A/c,k}^{\text{YM}}(p)\, p^2)\, \theta(k^2-p^2)$. 

\begin{figure}[t]
\begin{center}
  \includegraphics[width=.95\columnwidth]{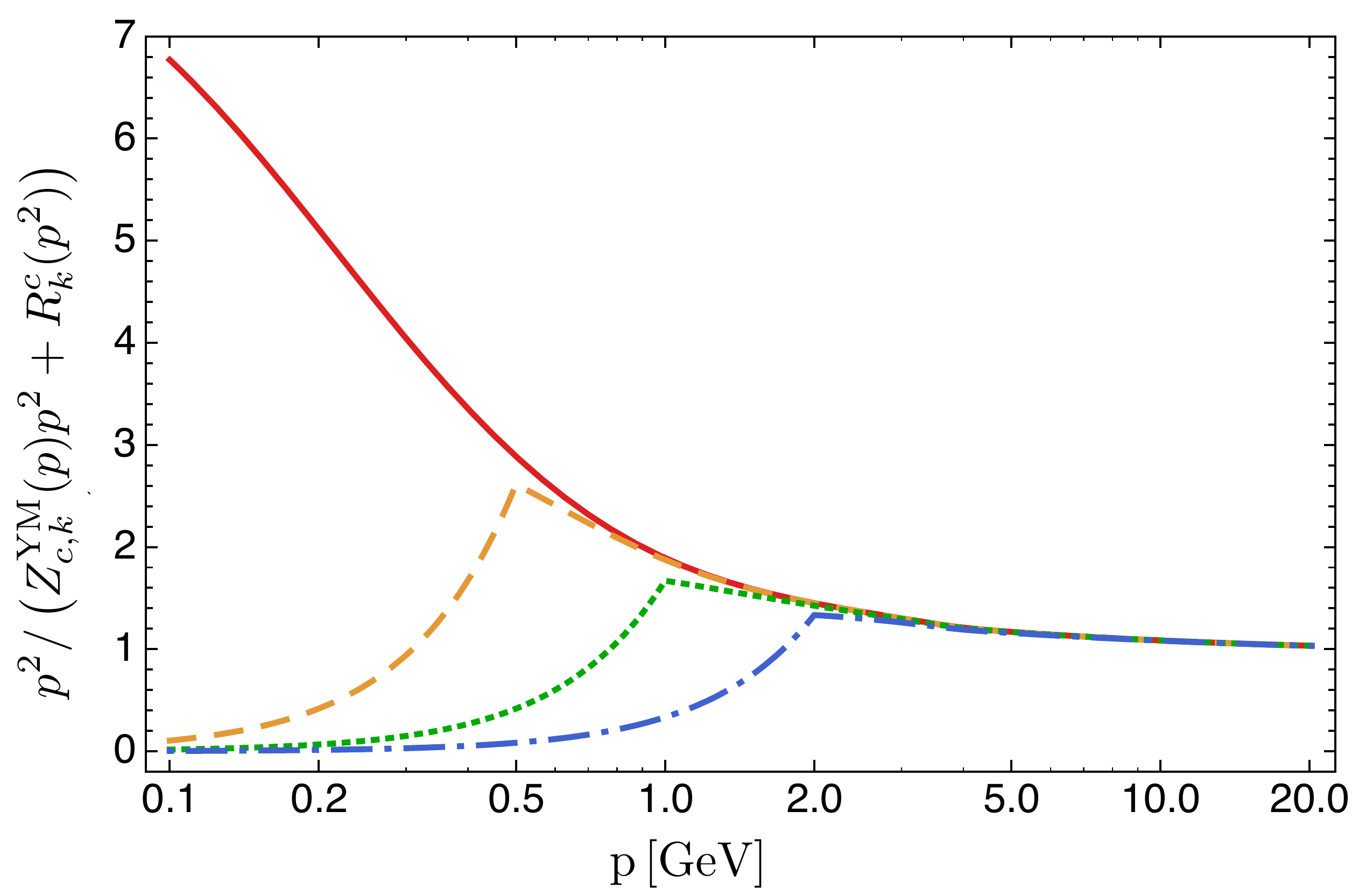}
  \caption{The regulated ghost dressing function from pure Yang-Mills theory as a function of the momentum for various $k$. We use this as an external input for our QCD computations. The labelling is the same as in Fig.~\ref{fig:gluein}.
  }\label{fig:ghin}
\end{center}
\end{figure}

We want to emphasise that a particular strength of the approach presented here is that it is independent of the specific form of the input in the sense that Yang-Mills propagators from any given method can be used. We have explicitly checked that our results are not altered if we use e.g. lattice input. In this case, the input dressing functions are of the form $Z_{A/c}^{\text{YM}}(p) = Z_{A/c,k=0}^{\text{YM}}(p)$ and the RG-scale dependence can be introduced by the identification $p\!=\!k$.

In order to make full use of the non-trivial
input we use here, we expand the flow equation for the gluon propagator in QCD
about that in Yang-Mills theory. We use the freedom in defining the
cutoff function $R_k^A$ to simplify the
analysis. This is done by choosing the same prefactor $Z_{A,k}$ for
the gluon regulator as for the vertex parametrisations in
\eq{eq:vertices}, see \Eq{eq:regdefs}.  Note that the gluon propagator enters in loop
integrals with momenta $p^2 \lesssim k^2$. If we estimate the full
gluon propagator \eq{eq:gluepropdef} with the simple expression (with
the tensor structure omitted for clarity)
\begin{align}\label{eq:simpleA}
  G_{A,k}(p) \approx \0{1}{Z_{A,k} \,p^2 +R_k^A} = \0{1}{Z_{A,k}
  }\,\0{1}{p^2 \left(1+r_B(p^2/k^2)\right)}\,,
\end{align} 
i.e.\ we only consider the fully $p$-dependent $Z_{A,k}(p)$ evaluated at
$p\!=\!k$, the system of flow equations considered is tremendously
simplified. The error of such a simple estimate relates to
\begin{align}\label{eq:Deltaglue}
 & p^3 \left(\0{1}{Z_{A,k}(p)\, p^2 +R_k^A} -
    \0{1}{Z_{A,k}\, p^2 +R_k^A}\right)^n\\ \nn
&\quad= p^{3+2n} \left(\0{Z_{A,k}-Z_{A,k}(p) }{
  \left(Z_{A,k}(p)\, p^2 +R_k^A\right) 
  \left(Z_{A,k}\, p^2 + R_k^A\right)} 
  \right)^n\,,
\end{align}
where the factor $p^3$ stems from the momentum integration $\sim dp\,p^3$.
The expression in \eq{eq:Deltaglue} occurs with powers $n\geq 1$ in
the difference of the full flow equations and the approximated flows
with \eq{eq:simpleA}, and is evaluated for momenta $p^2 \lesssim k^2$.
For small momenta it tends towards zero while its value for maximal
momenta $p^2 \approx k^2$ is proportional to the difference
$Z_{A,k}-Z_{A,k}(k)$. Consequently, we choose
\begin{align}\label{eq:ZAchoice} 
Z_{A,k}=Z_{A,k}(k)\,.
\end{align} 
We have checked that the difference between full flows and
approximated flows is less than 5\% for all $k$. \\
Within approximations \eq{eq:vertices} and \eq{eq:simpleA}, the gluon
propagator enters flow equations only via the anomalous
dimension $\eta_{A,k}$ with
\begin{align}\label{eq:etaA} 
\eta_{A,k} =-\0{\partial_t  Z_{A,k}}{Z_{A,k}}\,.
\end{align}
As a consequence of \eq{eq:ZAchoice}, $\eta_{A,k}$ has two
contributions from the full dressing function $Z_{A,k}(p)$,
\begin{align}\label{eq:etacontr}
  \partial_t Z_{A,k} = \partial_t Z_{A,k}(p)\bigr|_{p^2=k^2} + k
  \frac{\partial Z_{A,k}(p)}{\partial p}\biggr|_{p^2=k^2}\,.
\end{align}
The first term stems from the genuine $k$-dependence of the dressing
function, while the second term results from its momentum
dependence. As it is the case for any flow of a coupling in a gapped
theory (away from potential fixed points), the first term of
\eq{eq:etacontr} vanishes in the limit $k \!\rightarrow\! 0$,
\begin{align}
  \lim_{k\rightarrow 0} \partial_t Z_{A,k}(p)\bigr|_{p^2=k^2} =0\,.
\end{align}
The second term of \eq{eq:etacontr} carries the information about the
momentum dependence of the dressing function and in particular of the
(bare) mass gap $m_\text{gap}$ at small momenta. The gluon propagator
exhibits a gap at small momentum scales and hence the dressing
function of the full quantum theory, $Z_{A,k=0}(p)$, is of the form
\begin{align}\label{eq:etacontr1}
\lim_{p^2\rightarrow 0} Z_{A,k=0}(p) \propto \frac{m_\text{gap}^2}{p^2}\,.
\end{align}
This implies for the second term in \eq{eq:etacontr}
\begin{align}\label{eq:etacontr2}
  \lim_{k\rightarrow 0} k \partial_{p} \ln
  \left(Z_{A,k}(p)\right)\Bigr|_{p^2=k^2} = -2\,.
\end{align}
Thus, the second term of \eq{eq:etacontr} generates a non-vanishing
gluon anomalous dimension $\eta_{A,k}$, as defined in \eq{eq:etaA} for
$k \!\rightarrow\! 0$.

We note that this difference between the pure $k$-dependence and the
momentum dependence of the gluon dressing function is both highly
non-trivial and indispensable in any satisfactory truncation, even on
a qualitative level. The RG-scale dependence alone does not suffice to
capture the non-perturbative physics of YM theory or QCD in the gauge
sector, as it misses the confining properties of the theory. Being of
primary importance, the gluon mass gap emerges from the non-trivial
momentum dependence of the propagator. We remark that this is in
contrast to the chiral properties of the matter sector of QCD, where
approximations based on solely $k$-dependent parameters at least
qualitatively capture all the relevant physics.  

It is crucial that $Z_{A,k}$ does not appear explicitly, and hence
flows do only depend on $\eta_{A,k}$, the vertex couplings $g$, masses
and further couplings. Note that this is only partially due to the
approximation in \eq{eq:simpleA}. It mainly relates to the
parameterisations \eq{eq:vertices} of the vertices which stores most
of the non-trivial information in the associated vertex couplings
\begin{align}\label{eq:alphas}
 \alpha_{i}=\0{g_i^2}{4\pi}\,, \quad {\rm with}\quad i= {\bar c Ac}\,, 
    A^3\,, {A^4}\,, {\bar q Aq} \,.
\end{align}
This freedom directly relates to the reparametrisation invariance of the
theory and hence to RG invariance. The above discussion in particular
applies to the anomalous dimension itself: first, we note that the glue
part $\eta_{{\rm glue},k}$ of the anomalous dimension $\eta_{A,k}$
only depends on the vertex couplings:
\begin{align}\label{eq:etaglue}  
\eta_{{\rm glue},k}=\eta_{{\rm glue},k}(\alpha_{\bar c Ac}\,,  
\alpha_{A^3}\,,\alpha_{A^4})\,.  
\end{align}
In the semi-perturbative regime these couplings agree due to the
(RG-)modified Slavnov--Taylor identities
\cite{Ellwanger:1994iz,D'Attanasio:1996jd,Igarashi:2001mf,Pawlowski:2005xe}, which
themselves do not restrict the couplings in the non-perturbative
transition regime, see Ref.~\cite{Mitter:2014wpa}. In turn, in the
non-perturbative regime the couplings differ already due to their
different scalings with the gluonic dressing $Z_{A,k}$. For small
cutoff scales $k\to 0$, this dressing diverges proportional to the QCD
mass gap,
\begin{align}\label{eq:massgap}
\lim_{k\to 0}Z_{A,k} \propto \bar m_{\rm gap}^2 = \0{m^2_{\rm gap}}{k^2}\,.
\end{align}
This is a slight abuse of notation since $\bar
m^2_{\rm gap}$ in \eq{eq:massgap} is not renormalised as the other
dimensionless mass ratios $\bar m^2$. Here it simply relates to the
wave-function renormalisation $Z_{A,k}$ defined in
\eq{eq:ZAchoice}. Hence, it is not RG-invariant and should not be
confused with the physical mass gap of QCD. It is related with the
latter upon an appropriate renormalisation.

As a consequence, while we expect $\alpha_{\bar c Ac}\approx \alpha_{\bar q Aq}$
down to small scales, the purely gluonic couplings should be suppressed
to compensate the higher powers of diverging $Z_{A,k}$ present in the
vertex dressing in \eq{eq:vertices}. This also entails that we may
parameterise the right hand side with powers of $1/\alpha_i$. For $i =\bar c Ac,\, \bar q Aq$, for example, we expect
$1/\alpha_i$. In accordance with this observation, we parameterise the
difference of the various vertex couplings in $\eta_{\rm glue}$ with
the gap parameter $\bar m_{\rm gap}$ defined in \eq{eq:massgap} and
conclude for the gluon anomalous dimension of QCD
\begin{align}\label{eq:gprop}
  \eta_{A,k} = \eta_{{\rm
      glue},k}(\alpha_s,\bar m_{\rm gap}) + \Delta\eta^{\ }_{A,k}(\alpha_{\bar q A
    q}, \bar m_q)\,,
\end{align} 
where $\alpha_s$ stands for either $\alpha_{\bar c Ac}$ or
$\alpha_{A^3}$. 
We shall check that our results do not depend
on this choice which justifies the identification of the couplings in
\eq{eq:gprop}. Note that this does not entail that the couplings agree
but that they differ only in the regime where the glue
fluctuations decouple. Moreover, in the present approximation
$\alpha_{A^4}$ is not computed separately but identified with
$\alpha_{A^3}$.

A simple reduction of \eq{eq:gprop} is given by
\begin{align}\label{eq:eta-simple}
  \eta_{A,k} = \eta_{{A},k}^{\rm YM} + \Delta\eta^{\ }_{A,k}(\alpha_{\bar q A
    q}, \bar m_q)\,.  
\end{align} 
This amounts to a gluon propagator, where the vacuum polarisation is
simply added to the Yang-Mills propagator. This approximation has been
used in an earlier work, \cite{Braun:2008pi,Braun:2009gm,Pawlowski:2010ht}, and
subsequently in related Dyson-Schwinger works, see e.g.\
\cite{Fischer:2011mz,Fischer:2012vc,Fischer:2013eca,Fischer:2014vxa}.

\begin{figure}[t]
\begin{center}
  \includegraphics[width=.95\columnwidth]{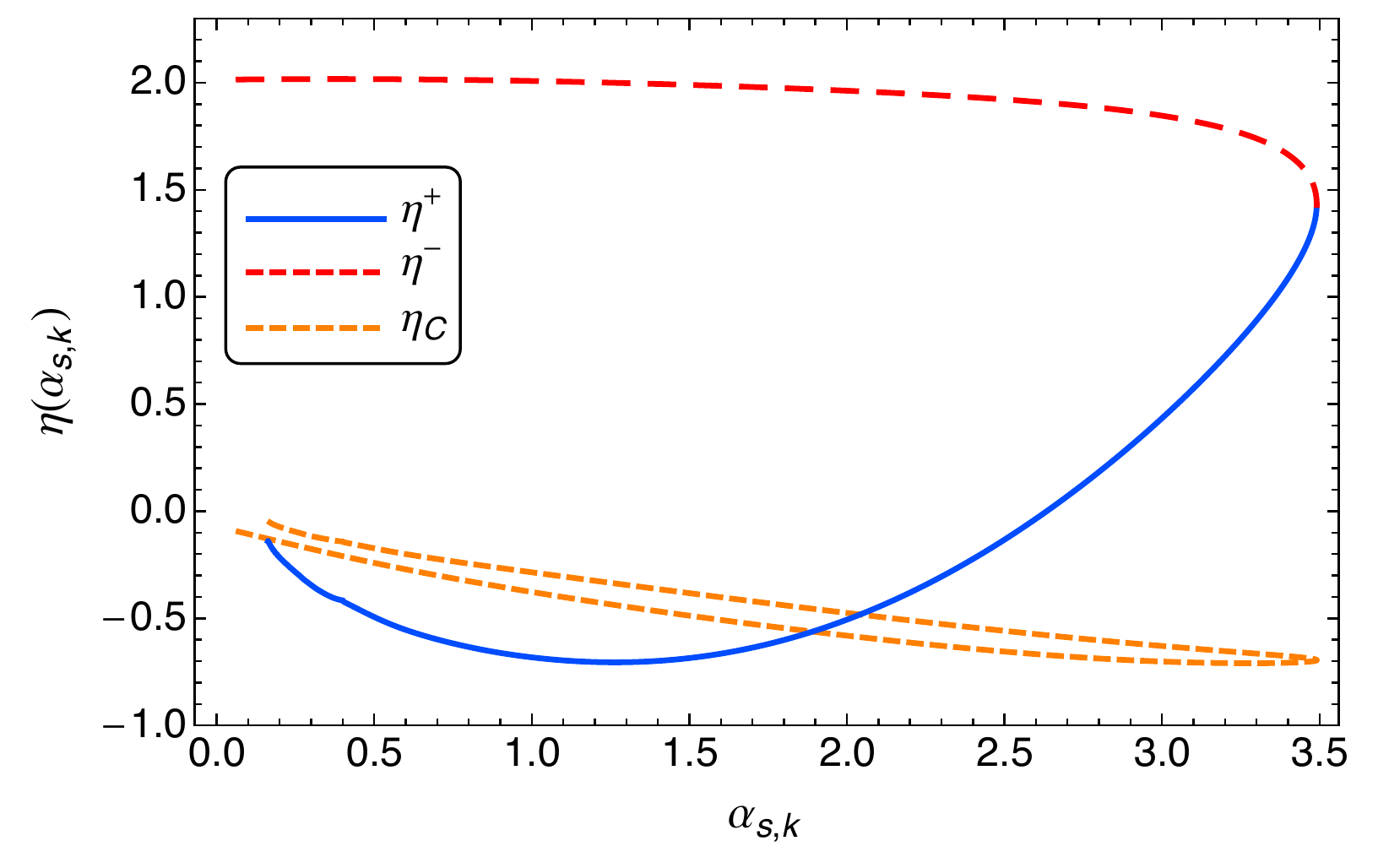}
  \caption{ The UV and IR branches of $\eta_A^{\rm YM}$, $\eta^{+}$
    and $\eta^{-}$, as a function of the strong coupling.
  }\label{fig:eta}
\end{center}
\end{figure}

The term $\Delta\eta_{A,k}$ is the quark contribution to the gluon
anomalous dimension, and is computed with 
\begin{align}\label{eq:Deltaq}
\includegraphics[height=8.8ex]{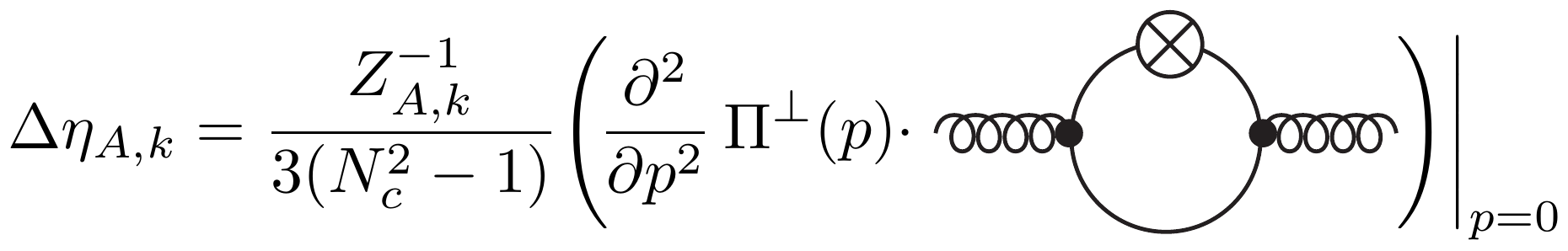}
\end{align}
Here, $p$ is the modulus of the external momentum and
$\Pi^\perp$ is the transversal projection operator defined in
\eq{eq:Pi}. Note that the dots represent full vertices and the
lines stand for full propagators. The crossed circle represents the
regulator insertion. For $N_f=2$ and $N_c=3$ we find 
\begin{align}\label{eq:gluevacpol}
\begin{split}
\Delta\eta_{A,k} =& \frac{1}{24\pi^2}\, g_{\bar{q} A q,k}^2 (1+\bar{m}_{q,k}^2)^{-4}\\
&\times\left[ 4-\eta_{q,k}+4 \bar{m}_{q,k}^2-(1-\eta_{q,k})\bar{m}_{q,k}^4 \right].
\end{split}
\end{align}
The approximation \eq{eq:Deltaq} works well as long as the quark
contribution has only a mild momentum dependence. This is the case due
to the gapping of the quarks via spontaneous chiral symmetry breaking,
and has been checked explicitly. A necessary check for the validity of this equation is that it reduces to the perturbative result in the corresponding limit, i.e. $\eta_{q,k},\,\bar{m}_{q,k} \rightarrow 0$. Indeed, \eq{eq:gluevacpol} reduces to one-loop perturbation theory in this case,
  $\Delta\eta_{A,k} = g_{\bar{q} A q,k}^2/(6\pi^2)$.

This leaves us with the task of determining $\eta_{{\rm glue},k}(\alpha_s,\bar m^2_{\rm gap})$, the pure glue contribution to
$\eta_{A,k}$. The loop expression for $\eta_{{\rm glue}}$ only
consists of Yang-Mills diagrams. As it depends solely on the value of the
coupling $\alpha_s$ we arrive at 
\begin{align} \label{eq:etaglue} 
\eta_{\rm glue}(\alpha_s,\bar
  m_{\rm gap}^{\rm QCD}) = \eta^{\rm YM}_{A}(\alpha_s,\bar m_{\rm
      gap}^{\rm QCD})\,,
\end{align}
i.e.\ the pure gauge part of the gluon anomalous dimension of QCD is
identical to the gluon anomalous dimension of pure Yang-Mills theory
however driven by the QCD couplings.  $\eta^{\rm YM}_A$ can be
determined in Yang-Mills theory or in quenched QCD as a function of
$\alpha_s$ and $\bar m_{\rm gap}$.

For using \eq{eq:etaglue}, a trackable form of $\eta^{\rm YM}_A$ as well
as $ \bar m_{\rm gap}^{\rm QCD}$ is required. To this end, we first note that $\eta(\alpha_{s,k})$ is a multi-valued
function in both Yang-Mills theory/quenched QCD and QCD, see
\Fig{fig:eta}. The two branches meet at $k=k_{\rm peak}$ (peak of the
coupling) with 
\begin{align}\label{eq:peak}
\left. \partial_t \alpha_{s,k}\right|_{k=k_{\rm peak}} =0\,. 
\end{align}

\begin{figure}[t]
\begin{center}
  \includegraphics[width=.95\columnwidth]{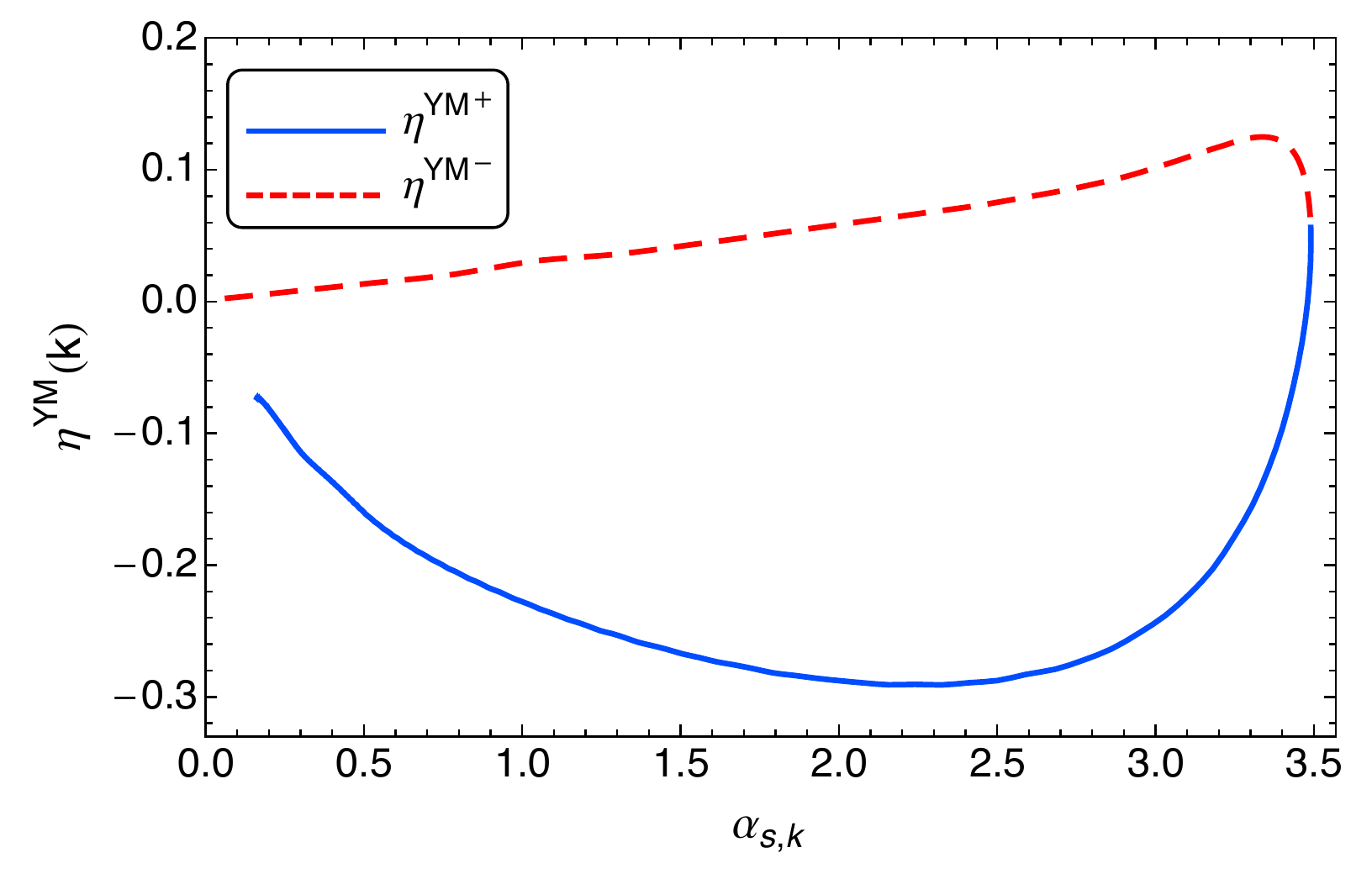}
  \caption{ The UV and IR branches of $\eta^{\rm YM}_{A,k}(k)$, which is defined in \eq{eq:fulleta}.
  }\label{fig:fulleta0}
\end{center}
\end{figure}

We have a UV branch $\eta^{+}(\alpha_s,\bar m_{\rm gap})$ for
$k>k_{\rm peak}$ and an IR branch $\eta^{-}(\alpha_s,\bar m_{\rm
  gap})$ for $k<k_{\rm peak}$. In \Fig{fig:eta} we show $\eta_A^{\rm
  YM}$ as a function of the coupling. Interestingly,
$\eta^+(\alpha_{s,k})$ is well-described by a quadratic fit in
$\alpha_s$ up to couplings close to $\alpha_{s,k_{\rm peak}}$. In
turn, $\eta^-(\alpha_{s,k})$ is well-described as a function of the
cutoff scale as indicated by \eq{eq:massgap}. In the deep IR the
  gluon dressing function is determined by the bare gap,
  $Z_{A,k\rightarrow 0} \propto m_\text{gap}^2/k^2$, see also the
  discussion around \eq{eq:etacontr2}. Hence we have
\begin{align}
\lim_{k\rightarrow 0} \eta_{A,k} = 2\,.
\end{align}
This is seen in \Fig{fig:eta}. We also see in this figure that the
whole IR branch $\eta^-$ is almost constant. This implies that the
mass gap which suppresses $\alpha_{s,k}$ develops quickly around
$k\!\approx\!k_\text{peak}$ and remains roughly constant for the rest
of the flow for $k \lesssim k_\text{peak}$. This allows us to
parametrise the IR-branch in terms of the RG-scale,
\begin{align}
\eta^-= 2 - c^- k^2\,,\quad {\rm with}\quad c^- = \0{2- 
\eta_A^{\rm YM}(\alpha_{\rm peak})}{k_{\rm peak}^2} \,, 
\end{align}
where the mass gap $\bar m^2_{\rm gap}$ relates to $\eta_A^{\rm
  YM}(\alpha_{\rm peak})$. Note that the quality of these simple fits
entails that the transition from the semi-perturbative regime to the
non-perturbative IR regime happens quite rapidly and asymptotic fits in
both areas work very well. In summary we arrive at the final representation of
$\eta^{\rm YM}_A$ with
\begin{align}\label{eq:repeta}
\begin{split}
  \eta_{A,k}^{\rm YM}(\alpha_{s,k}) =&\eta^+(\alpha_{s,k})\theta(\alpha_{s,k}-\alpha_{s,\rm
    peak})\\
& +\eta^-(k)\theta(\alpha_{s,\rm
    peak}-\alpha_{s,k})\,.
\end{split}
\end{align} 
Inserting \eq{eq:repeta} on the right hand side of \eq{eq:etaglue}
gives us a closed equation for $\eta_{A,k}$ in \eq{eq:gprop}. Its
integration also provides us with the QCD mass gap.

The same analysis as for $\eta_{A,\,k}$ can be applied to the ghost
anomalous dimension $\eta_{c,\,k}$ leading to a similar representation with
the only difference that $\eta_{c,k=0} =0$. It turns out that an even
simpler global linear fit gives quantitatively reliable results for
matter correlations,
\begin{align}\label{eq:ghprop}
  \eta_{c,k}(\alpha_{s,k})= \0{ \alpha_{s,k}}{ \alpha}
    \eta_{c,k}^{\rm YM }(\alpha)\,,
\end{align} 
where $\alpha^{\ }_{s,k}= \alpha^{\ }_{\bar c Ac,k}$, see \Fig{fig:eta}. This
modification is used in the equation for the ghost-gluon vertex. Note
that this overestimates ghost-gluon correlations in the deep infrared
where the glue-sector has decoupled from the matter sector. Hence this
is of no relevance for the physics of chiral symmetry breaking
discussed in the present work. 

We are now in a position to finally determine the ghost and gluon
propagators at vanishing cutoff scale in dynamical QCD. Again, we could
use the $\alpha, \bar m_{\rm gap}$ representation for extracting the
full dressing function $Z_{A,k}(p)$ on the basis of the results. To
that end, the momentum-dependent flows as functions of $\alpha, \bar
m_{\rm gap}$ are required,
\begin{align}\label{eq:fulleta} 
\eta^{\rm YM}_{A,k} (p) = -\0{\partial_t Z^{\rm YM}_{A,k}(p)}{Z^{\rm YM}_{A,k}(p)}\,,\qquad 
\partial_t \Delta\eta_{A,k}(p)\,, 
\end{align}
where $\Delta\eta_{A,k}(p)$ stands for the momentum-dependent flow of
the vacuum polarisation. The first term in \eq{eq:fulleta} again is
well approximated in terms of a low order polynomial in
$\alpha_s$. This is expected because is relates directly to the
standard anomalous dimension of the gluon. In \Fig{fig:fulleta0} it is
shown for momentum $p=k$ as a function of $\alpha_{s,k}$. The
definition of $\eta^{\rm YM}_{A,k} (p)$ implies that only the first
term in \eq{eq:etacontr} contributes here. Thus, for vanishing $k$
\eq{eq:etacontr1} holds and hence $\lim_{k\rightarrow 0}\eta^{\rm
  YM}_{A,k} (k) = 0$ as observed in \Fig{fig:fulleta0}.

\begin{figure}[t]
\begin{center}
  \includegraphics[width=.95\columnwidth]{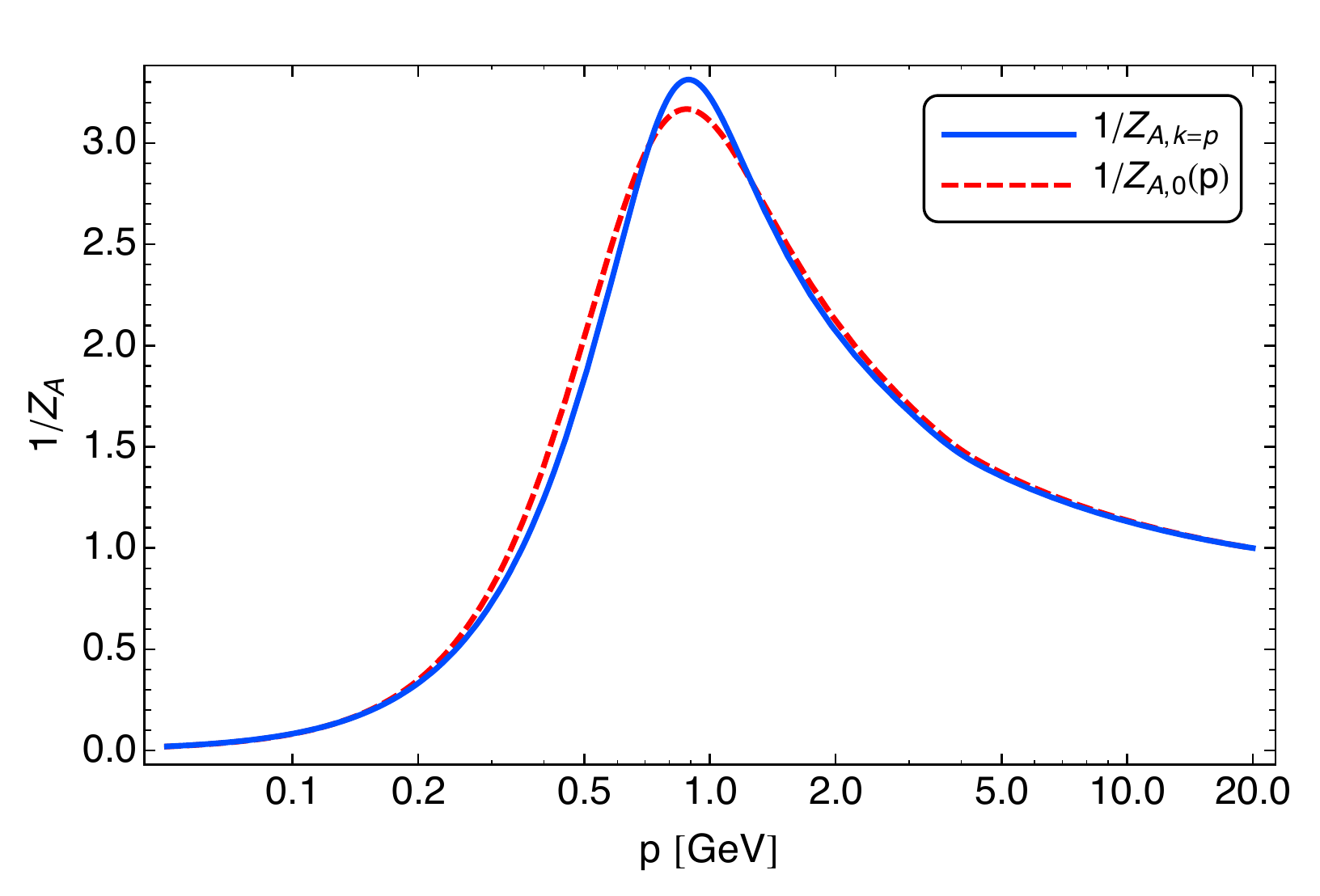}
  \caption{Comparison of the momentum dependent gluon dressing
    function $Z_{A,0}(p)$ and $Z_{A,k=p}$.}\label{fig:zkzp}
\end{center}
\end{figure}

An already very good estimate for the
dressing function is
\begin{align}\label{eq:approxZ1} 
Z_{A,k=0}(p) \simeq Z_{A,k=p}(p) = Z_{A,k=p}\,,
\end{align}   
as the flow of the propagators decay rapidly for momenta larger than
the cutoff scale, $p\gtrsim k$. Moreover, the momentum derivative of
the dressing is only large in the UV-IR transition regime. In
\Fig{fig:zkzp}, the inverse dressing $1/Z_{A,0}(p)$ and its
approximation $1/Z_{A,p}$ are shown. Clearly, there are only minor
deviations in the UV-IR transition regime. The same argument holds
true to an even better degree for the quark contribution, and we have
checked the smoothness of the flow $\Delta\Gamma_{A,k}(p)$. This leads
to a very simple, but quantitative estimate for the full dressing
function with
\begin{align} \label{eq:glueprop} Z^{\rm glue}_{A/c,k=0}(p) \simeq
  \0{ Z^{\rm YM}_{A/c,k=0}(k_\alpha)}{Z^{\rm
      YM}_{A/c,k_\alpha}}\, Z^{\rm glue}_{A/c ,k=p}\,,
\end{align}
with 
\begin{align}  \label{eq:Zeta}
  Z^{\rm glue}_{A/c,k}  = \exp\left\{ -\int_\Lambda^p \0{dk}{k} \eta^{\rm glue}_{A/c,k} \right\}\,,
\end{align}
where $Z_{A/c,\Lambda}=1$, and $k_\alpha=k(\alpha_{s,k})$ is the YM-cutoff
value that belongs to a given coupling $\alpha_s$.  

In summary we conclude that, based on \Fig{fig:zkzp}, an already
quantitative approximation to the fully unquenched propagator is done if 
putting the ratio in \eq{eq:glueprop} to unity. This leads to  
\begin{align}
  Z_{A/c}(p) \simeq \exp\left\{ -\int_\Lambda^p \0{dk}{k} \eta_{A/c,k} \right\}\,,
\end{align}
with $\eta_{A/c,k}$ defined in \eq{eq:gprop}. 
In the non-perturbative regime diagrams involving an internal gluon
are suppressed with the generated gluon mass. Hence, albeit the
approximation by itself may get less quantitative in the infrared, the
error propagation in the computation is small. 

In summary this leaves us with relatively simple analytic flow
equations for the fully back-coupled unquenching effects of glue and
ghost propagators. A full error analysis of the analytic
approximations here will be published elsewhere, and is very important
for the reliable application of the present procedure to finite
temperature and density.

In the following, we will outline the definition and derivation of the
gluonic vertices we use. First of all, we only take into account the
classical tensor structure of the vertices. Moreover, throughout this
work, we define the running coupling at vanishing external momentum.
Together with our choice for the regulators, this has the advantage
that the flow equations are analytical equations. In particular,
loop-momentum integrations can be performed analytically. This
approximation is semi-qunatitative as long as the dressing of the
classical tensor structures do not show a significant momentum
dependence, and the other tensor structures are suppressed. 

This approximation is motivated by results on purely gluonic vertices,
see
Refs.~\cite{Fister:2011uw,Fister:Diss,Huber:2012kd,Pelaez:2013cpa,%
  Eichmann:2014xya,Blum:2014gna,Binosi:2014kka,Gracey:2014mpa,Cyrol:2014kca},
which show non-trivial momentum-dependencies only in momentum region
where the gluon sector already starts to decouple from the system. In
turn, the tensor structures and momentum dependences of the
quark-gluon vertex are important, see the DSE studies
\cite{Hopfer:2013np,Williams:2014iea,Aguilar:2014lha} and the recent
fully quantitative FRG study \cite{Mitter:2014wpa}. To take this
effectively into account, we introduce an infrared-strength function
for the strong couplings, which is discussed at the end of this
section and in App.~\ref{app:IRstrength}.

To extract the flow of the quark-gluon coupling $g_{\bar q A q}$, we
use the following projection procedure,
\begin{align}
\begin{split}
  \partial_t g_{\bar q A q}&=\frac{1}{8 N_f (N_c^2-1)}\\
  &\quad\times \lim_{p\rightarrow 0}\left.\text{Tr}\left(\gamma_\mu
      t^a\frac{\partial_t \Gamma_{k}}{\delta q\delta A^a_\mu\delta\bar
        q}\right)\right|_{\Phi=\Phi_0},
\end{split}
\end{align}
which leads to the equation
\begin{align}\label{eq:quarkgluonflow}
\begin{split}
  &\partial_t g_{\bar q A q,k} =\\
  &\quad\frac{1}{2}\left(\eta_{A,k}+2\eta_{q,k} \right) g_{\bar q A q,k}\\
  &\quad-v(d)\,g_{\bar q A q,k}\,\bar{h}_k^2\left\{ \CN_{2,1}^{(m)}(\bar{m}_{q,k}^2,
    \bar{m}_{\sigma,k}^2;\eta_{q,k},\eta_{\phi,k})\right.\\
  &\quad + \left.(\Nf^2-1)\,\CN_{2,1}^{(m)}(\bar{m}_{q,k}^2,
    \bar{m}_{\pi,k}^2;\eta_{q,k},\eta_{\phi,k}) \right\}\\
  &\quad +g_{\bar q A q,k}^3\frac{12
    v(d)}{\Nc}\,\CN_{2,1}^{(g)}(\bar{m}_{q,k}^2;\eta_{q,k},
  \eta_{A,k})\\
  &\quad + g_{\bar q A q,k}^2\, g_{A^3,k} \,3 v(d)
  \Nc\,\CN_{1,2}^{(g)}(\bar{m}_{q,k}^2;\eta_{q,k},\eta_{A,k})\,.
\end{split}
\end{align}
The threshold functions appearing on the right-hand side can be
found in the App.~\ref{app:thres}. For the quark-gluon vertex, no
ghost diagrams are present. Furthermore, the mesonic contributions
dominate in the infrared. These contributions have the same sign as
the gluonic ones and therefore lead to an effective infrared
enhancement of the quark-gluon vertex. The three-gluon vertex $g_{A^3,k}$ is defined via 
\begin{align}
 \partial_t g_{A^3,k}&=\frac{i}{12 N_c (N_c^2-1)}\lim_{p\rightarrow
    0}\frac{\partial^2}{\partial p^2}\\ \nonumber
  &\quad\left.\text{Tr}\left(\delta_{\mu\nu}p_\sigma f^{abc}\frac{
        \partial_t\Gamma_{k}}{\delta A(p)_\mu^a\delta
        A(-p)_\nu^b\delta A_\sigma^c(0)}\right)\right|_{\Phi=\Phi_0}\,.
\end{align}
Note that in the limit of vanishing external momentum the flow is
independent of the kinematic configuration in the projection
procedure.  Thus, we find for the flow equation for $N_c=3$ and
$N_f=2$
\begin{align}\label{eq:gaaa}
  \begin{split}
    \partial_t g_{A^3,k} &= \frac{3}{2} \eta_{A,k}\, g_{A^3,k}\\
    &\quad -\frac{1}{6 \pi^2}\, g_{\bar q A
      q,k}^3\left(1-\frac{\eta_{q,k}}{4}\right)
    \frac{(1+2\bar m_{q,k}^2)}{(1+2\bar m_{q,k}^2)^4}\\
    &\quad+\frac{3}{64\pi^2}\, g_{A^3,k}^3 \left(11-2\eta_A\right)\\
    &\quad+\frac{1}{64\pi^2}\, g_{\bar c A c,k}^3 \left(
      1-\frac{\eta_{C,k}}{8} \right)\,,
\end{split}
\end{align}
with the ghost anomalous dimension $\eta_{C,k}=-(\partial_t
Z_{C,k}(k^2))/Z_{C,k}(k^2)$. The second line in \eq{eq:gaaa}
corresponds to the quark-triangle diagram and the third and fourth
line are the gluon- and ghost-triangle diagrams, respectively. Note
that the third line also includes the contribution from the diagram
containing the four-gluon vertex, which we approximate as explained
below.

\begin{figure}[t]
\begin{center}
  \includegraphics[width=1.\columnwidth]{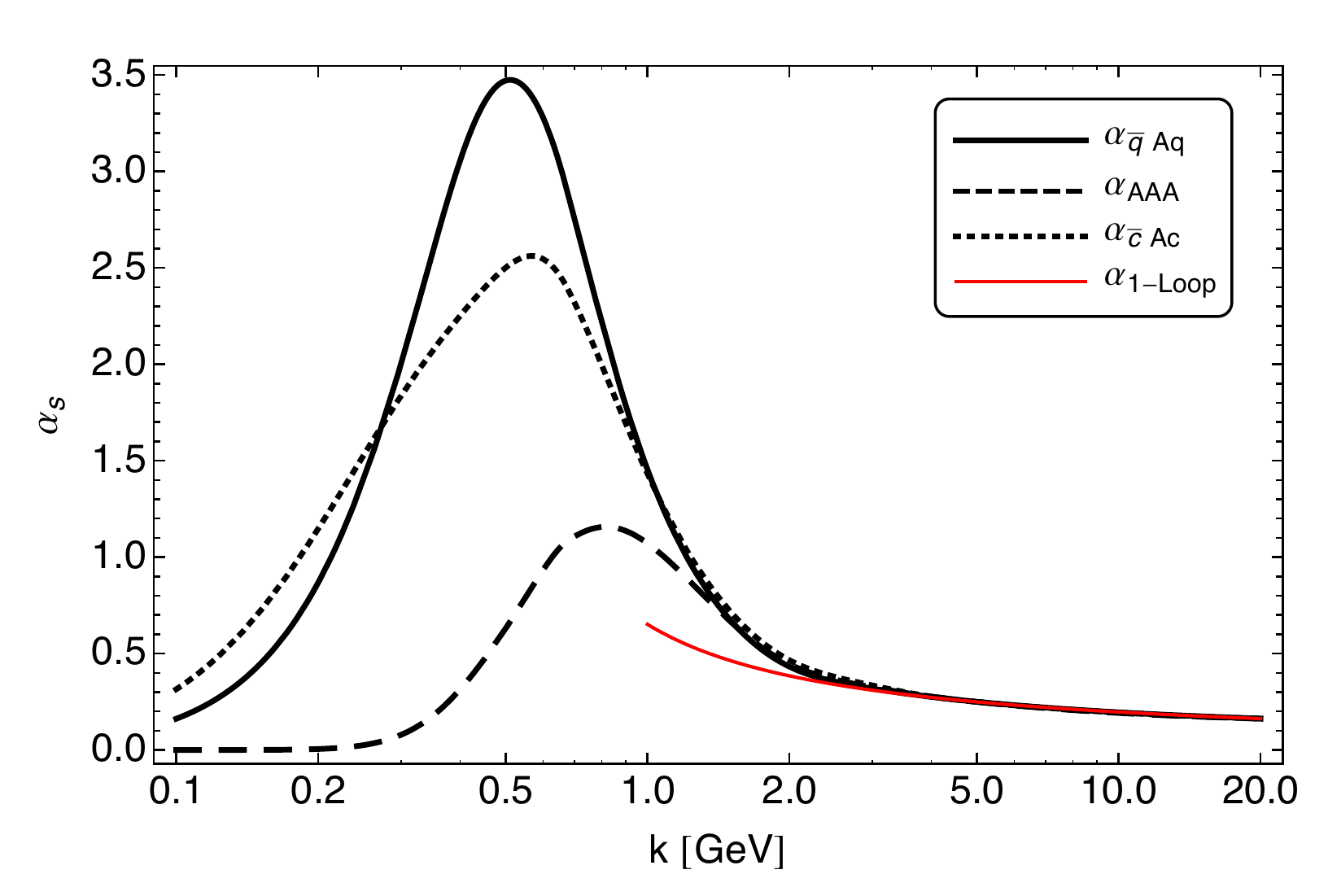}
  \caption{The running of the different strong couplings in comparison
    to the 1-loop running. Since perturbation theory breaks down at the scale where the strong couplings start to deviate from each other, we show the 1-loop running only down to $1\,\text{GeV}$.}\label{fig:alphas}
\end{center}
\end{figure}

Within our approximation, the ghost-gluon vertex $g_{\bar c A c,k}$
has only canonical running since the diagrams that contribute to the
flow of $g_{\bar c A c,k}$ are proportional to the external
momentum. Thus, at vanishing external momentum they vanish and we are
left with:
\begin{align}\label{eq:gcac}
  \partial_t g_{\bar c A c,k} &=
  \left(\frac{1}{2}\eta_{A,k}+\eta_{C,k}\right) g_{\bar c A c,k}.
\end{align}
Lastly, we comment on our approximation for the four-gluon vertex
$g_{A^4,k}$. For the sake of simplicity, we restrict here to a
semi-perturbative ansatz for this vertex, which ensures that
$g_{A^4,k}$ has the correct perturbative running. To this end, we set
\begin{align}
g^2_{A^4,k} = g_{A^3,k}^2\,.
\end{align}
This approximation is valid for $k\gtrsim 1.5~\text{GeV}$. For smaller
scales, non-perturbative effects potentially lead to a different
running.

This leads to an explicit expression for $\Delta\Gamma_\text{glue}$ in
\eq{eq:trunc}:
\begin{align}\label{eq:deltagammaglue}
\begin{split}
  \Delta\Gamma_\text{glue} &= \int_x\biggl\{\frac{1}{4}\left(F^2
    \Bigr|_{g_k=g_{A^3,k}} - F^2\Bigr|_{g_k}\right)\\
  &\quad + \bar c^a\partial_\mu\left(D_\mu^{ab}
    \Bigr|_{g_k=g_{\bar c A c,k}} - D_\mu^{ab}\Bigr|_{g_k}  \right) c^b\\
  &\quad + Z_{q,k} \,\bar q\, \gamma_\mu\left(
    D_\mu\Bigr|_{g_k=g_{\bar q A q,k}} - D_\mu\Bigr|_{g_k} \right) q
  \biggr\}\,,
\end{split}
\end{align}
where we used the abbreviation $F^2 = F_{\mu\nu}^a F_{\mu\nu}^a$. We
see that $\Delta\Gamma_\text{glue}$ corrects for distinctive coupling
strengths for interaction terms. While perturbation theory ensures
that all couplings agree in the UV, non-perturbative effects lead to
differing behaviour in the mid-momentum and IR regime.

The result for the different running couplings discussed here is shown
in \Fig{fig:alphas}. While they all agree with each other and follow
the perturbative running at scales $k\gtrsim 3~\text{GeV}$,
non-perturbative effects induce different runnings at lower scales. In
particular, the former statement is a highly non-trivial consistency
check of the approximation we make here. 

As discussed above, in the present study we focus on the RG flows of
the most relevant couplings from a phenomenological point of view. In
particular, we concentrate on the effects of fluctuations on the
relevant and marginal parameters of the classical gauge action in
\eq{eq:trunc}. Consequently, non-classical interactions which are
potentially relevant are not taken into account here. Furthermore, we
only consider vertices at vanishing external momenta, although
momentum dependencies may play an important quantitative role. As an
example, this becomes apparent in the flow of the ghost-gluon vertex
\eq{eq:gcac}: while the diagrams driving the flow of $g_{\bar c A
  c,k}$ vanish within our approximation, they give finite
contributions at non-vanishing momenta. This was studied in more
detail in the case of quenched QCD \cite{Mitter:2014wpa}. Indeed, it turned out
that both, momentum dependencies and the inclusion of non-classical
vertices, lead to large quantitative effects. It was shown there that
within an extended truncation the approach put forward in the present
work leads to excellent quantitative agreement with lattice QCD
studies. 

We take the findings in \cite{Mitter:2014wpa} as a guideline for a phenomenological
modification of the gauge couplings. Effectively this provides
additional infrared strength to the gauge couplings in the
non-perturbative regime with $k \lesssim 2$ GeV. This additional
strength is adjusted with the current quark mass at vanishing
momentum.  This is reminiscent to similar procedures within
Dyson-Schwinger studies, see e.g.
\cite{Roberts:2000aa,Fischer:2006ub}, the details are given in
App.~\ref{app:IRstrength}.

\section{Results}\label{sec:results}

\begin{figure}[t]
\begin{center}
  \includegraphics[width=1.\columnwidth]{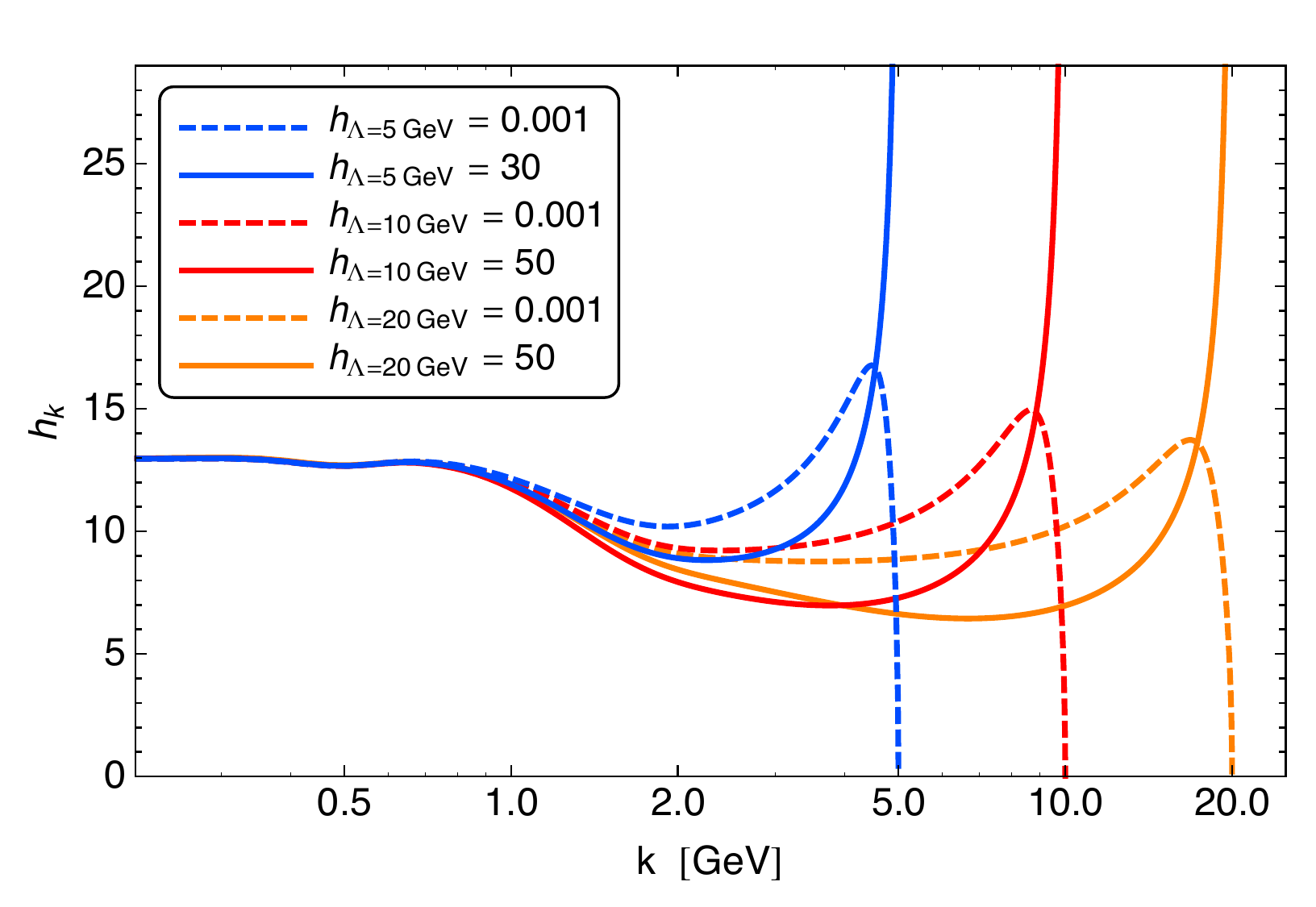}
  \caption{Yukawa coupling as a function of the RG scale for various
    initial scales $\Lambda$ and initial conditions
    $h_\Lambda$.}\label{fig:yukawa}
\end{center}
\end{figure}

First we summarize the system of flow equations used in the
  present work. The effective potential $\bar V_k(\bar \rho)$ and the
  Yukawa coupling $\bar h_k(\bar\rho)$ are expanded about a fixed bare
  field as shown in \eq{potential}. These expansions are already fully
  converged for $N_V\!=\!5$ and $N_h\!=\!3$, for a detailed discussion see
  \cite{Pawlowski:2014zaa}. The flow equations for the effective
  potential and its expansion coefficients are given by \eq{Vflow} and
  \eq{Vnflows}. For the Yukawa coupling they are given by
  \eq{flowYukawa2} and \eq{eq:hnflow} in the case of scale-independent
  meson fields. The latter are modified by dynamical hadronisation
  which results in \eq{eq:hadh1} for the final flow of the Yukawa
  coupling. The flows of the renormalised expansion point
  $\bar\kappa_k$ and the explicit symmetry breaking $\bar c_k$ are
  purely canonical and given by \eq{eq:kandc}. In order to accurately
  capture the physics in the IR, we choose the expansion point such
  that it matches the minimum of the renormalised effective potential
  at $k\!=\!0$, $\bar\kappa_{k=0} = \bar\rho_{0,k=0}$,
  cf. \cite{Pawlowski:2014zaa} for details. Owing to dynamical
  hadronisation, the flow of the four quark interaction
  $\bar\lambda_{q,k}$ for scale-independent fields enters through the
  flow of the Yukawa coupling and is given by
  \eq{eq:dtbarlaq}. Following our construction discussed in
  Sect.~\ref{sec:hadflows}, the flow of $\bar\lambda_{q,k}$ vanishes
  in the presence of the scale-dependent mesons. The RG flows of the
  quark-gluon, the three-gluon and the ghost-gluon couplings are given
  by \eq{eq:quarkgluonflow}, \eq{eq:gaaa} and \eq{eq:gcac}. Owing to
  our construction of the vertices, see \eq{eq:vertices}, non-trivial
  momentum dependencies of the propagators enter solely through the
  corresponding anomalous dimensions $\eta_{\Phi,k}$. For the mesons
  and quarks they are given by \eq{etasigma} and \eq{etaq}. The
  parametrisation of the gluon and ghost anomalous dimensions is
  discussed in section \ref{sec:gauge}. The gluon anomalous dimension
  $\eta_{A,k}$ is defined by \eq{eq:gprop} and contains the pure gauge
  part and the vacuum polarization. The vacuum polarization is given
  by \eq{eq:gluevacpol}. The pure gauge part is constructed from the
  full gluon anomalous dimension of pure Yang-Mills theory, which we
  use as an input. It is computed from \eq{eq:etaglue} and
  \eq{eq:repeta} with $\alpha_{s,k}\!=\!\alpha_{\bar c A c,k}$. The
  ghost anomalous dimension of QCD is computed from \eq{eq:ghprop},
  where we also augment the input from pure Yang-Mills theory by
  correcting for the differences between the strong couplings of YM
  and QCD, which, in turn, are computed here. Together with the fact
  that we evaluate all flows at vanishing external momentum, this
  leads to a set of ordinary differential equations in the RG scale
  $k$ which can easily be solved.

\begin{figure}[t]
\begin{center}
  \includegraphics[width=1.\columnwidth]{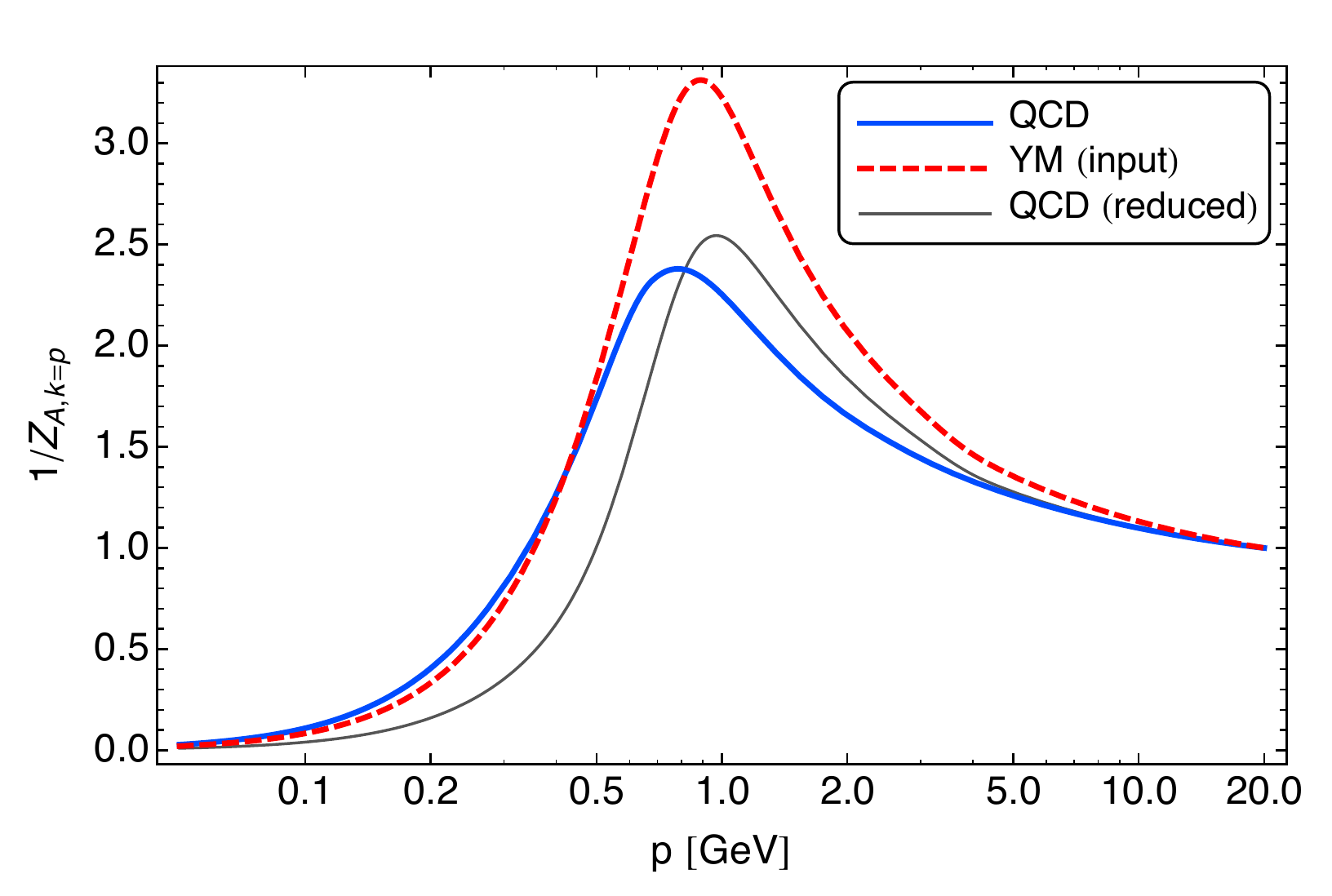}
  \caption{Comparison between the quenched and the unquenched running
    gluon propagators $1/Z_{A,k}^\text{YM}(k^2)$ and $1/Z_{A,k}(k^2)$
    as defined in Eq.~\eq{eq:glueprop}.  We also show the curve for QCD
    (reduced) where the gluon propagator is a direct sum of Yang-Mills propagator and 
vacuum polarisation, see Eq.~\eq{eq:eta-simple}. }\label{fig:Gprops}
\end{center}
\end{figure}

The starting point of the present analysis is the microscopic action
of QCD. We therefore initiate the RG flow at large scales, deep in the
perturbative regime. The initial values for the strong couplings are
fixed by the value of the strong coupling obtained from 1-loop
perturbation theory. Since the different strong couplings we use here
(see Eq.~\eq{eq:alphas}) need to be identical in the perturbative
regime, they consequently have the same initial value $\alpha_s$.  It
is shown in \Fig{fig:alphas} that indeed the different strong
couplings agree to a high degree of accuracy with the 1-loop running
of the strong coupling for scales $k>3\,\text{GeV}$. This is a very
important benchmark for the consistency of the approximations we
use. Note that the value of $\alpha_s$ implicitly determines the
absolute physical scale. Here we choose $\alpha_{s,\Lambda}=0.163$,
which relates to $\Lambda\approx 20\,\text{GeV}$. A quantitative
determination requires the determination of the RG-condition in
relation to standard ones such as the $\overline{\text{MS}}$-scheme as
well as the extraction of $\alpha_{s,k=0}(p=\Lambda)$, using $\Lambda$
as the renormalisation point. This goes beyond the scope of the
present paper and we shall restrict ourselves to observables that are
ratios of scales, our absolute scales are determined in terms of
$\Lambda=20$ GeV. The other microscopic parameter of QCD, the current
quark mass, is in our case fixed by fixing the symmetry breaking
parameter $c$. We choose $\bar c_\Lambda = 3.6\, \text{GeV}^3$ which
yields a infrared pion mass of $M_{\pi,0} = 137\, \text{MeV}$; $M_{k}
= k\bar m_{k}$ is the renormalized dimensionful mass.

\begin{figure}[t]
\begin{center}
  \includegraphics[width=.95\columnwidth]{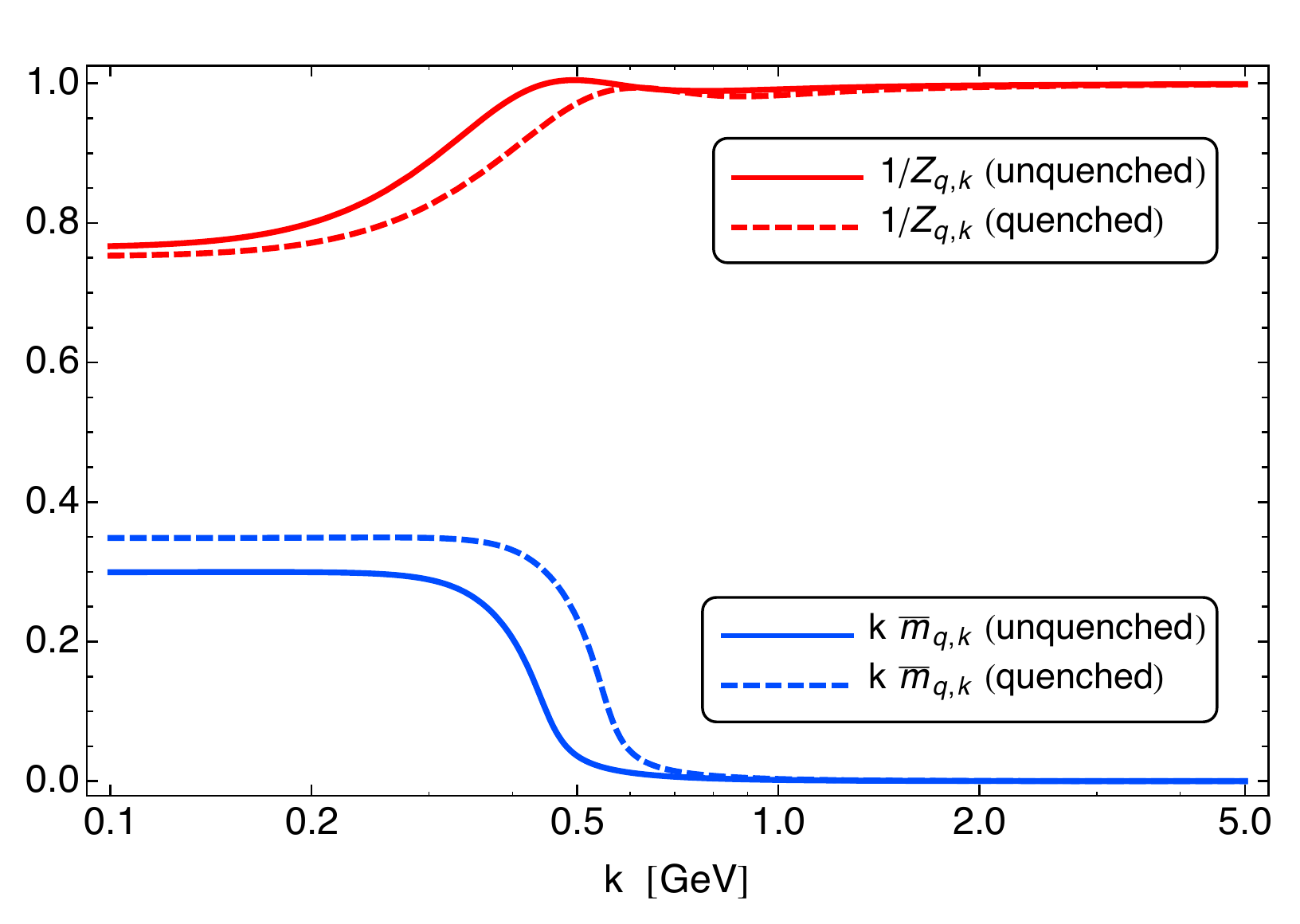}
  \caption{Dressing function (red) and mass (blue) of the quark as
    function of the RG scale at vanishing momentum. We compare our
    present model (solid) to the quenched model (dashed) with the
    parameters fixed to match those of
    \cite{Mitter:2014wpa}.}\label{fig:qprop}
\end{center}
\end{figure}

Note that the masses defined in Eq.~(\ref{eq:masses}), and
  hence in particular $M_{\pi,0}$, are curvature masses, i.e. the
  Euclidean two-point functions evaluated at vanishing
  momentum. However, it is the pole masses, defined via the poles of
  the propagators, that are measured in the experiments. Moreover,
  curvature and pole masses do not necessarily agree. In the present
  work, this difference is potentially of importance for the accurate
  determination of the pion mass. Now we use that curvature and pole
  masses are close for weakly momentum dependent wave function
  renormalisations, for a detailed discussion see
  \cite{Helmboldt:2014iya}. There it also has been shown that the
  pion wave function renormalisation is indeed weakly momentum
  dependent, and pion curvature and pole mass deviate by less than
  1\%. It has been also shown in \cite{Helmboldt:2014iya} that the large
  deviation of pion pole and curvature masses seen in previous works,
  \cite{Strodthoff:2011tz}, originates in the local potential
  approximation (LPA). Moreover, a scale-dependent, but
  momentum-independent, wave function renormalisation already removes
  the discrepancies seen in LPA, and the results agree well within the
  1\% level. In summary, curvature and pole mass of the pion agree on the 1\%
  level. The inclusion of momentum-independent running wave function
  renormalisations, as in the present work, guarantees quantitative
  reliability for this issue.

Since mesons are not present in the perturbative regime, we only have
to make sure that this sector is decoupled at the initial scale. We
therefore choose $M_{\pi,\Lambda}^2=M_{\sigma,\Lambda}^2 = 10^4
\Lambda^2$. Our results are independent of the choice of the initial
masses and the Yukawa coupling as long as the initial four-fermi
coupling related to it is far smaller than $\alpha_s^2$. This is
demonstrated for the Yukawa coupling in \Fig{fig:yukawa}, where we see
that, with initial values that differ by many orders of magnitude, we
always get the same solution in the IR. Loosely speaking, the memory
of the initial conditions is lost in the RG flow towards the IR regime
due to the presence to a pseudo fixed-point on intermediate scales,
see also Ref.~\cite{Gies:2002hq}.

In the present work we have studied the unquenching effects due to the
full back-coupling of the matter dynamics to the glue sector. In an
earlier work,\cite{Braun:2009gm,Pawlowski:2010ht}, we directly
identified $\eta_{{\rm glue},k}=\eta_{A,k}^{\rm YM}$ at the same
cutoff scale $k$, see Eq.~\eq{eq:eta-simple}.  This simply adds the
vacuum polarisation to the Yang-Mills propagator without feedback. It
is well-adapted for taking into account qualitatively even relatively
large matter contributions to the gluonic flow: the main effect of the
matter back-coupling is the modification of scales, most importantly
$\Lambda_{\rm QCD}$, which is already captured well in (one-loop)
perturbation theory, if the initial scale is not chosen too
large. This approximation has also been subsequently used in related
Dyson-Schwinger works, see e.g.\
\cite{Fischer:2011mz,Fischer:2012vc,Fischer:2013eca,Fischer:2014vxa},
extending the analysis also to finite density. Here, we improve these
approximations by taking into account the back-reaction of matter
fluctuations on the pure gauge sector. Furthermore, the gluon vacuum
polarization was based on a one-loop improved approximation in
previous FRG studies. Here, we compute the full vacuum polarization
self-consistently.

\begin{figure}[t]
\begin{center}
  \includegraphics[width=1.\columnwidth]{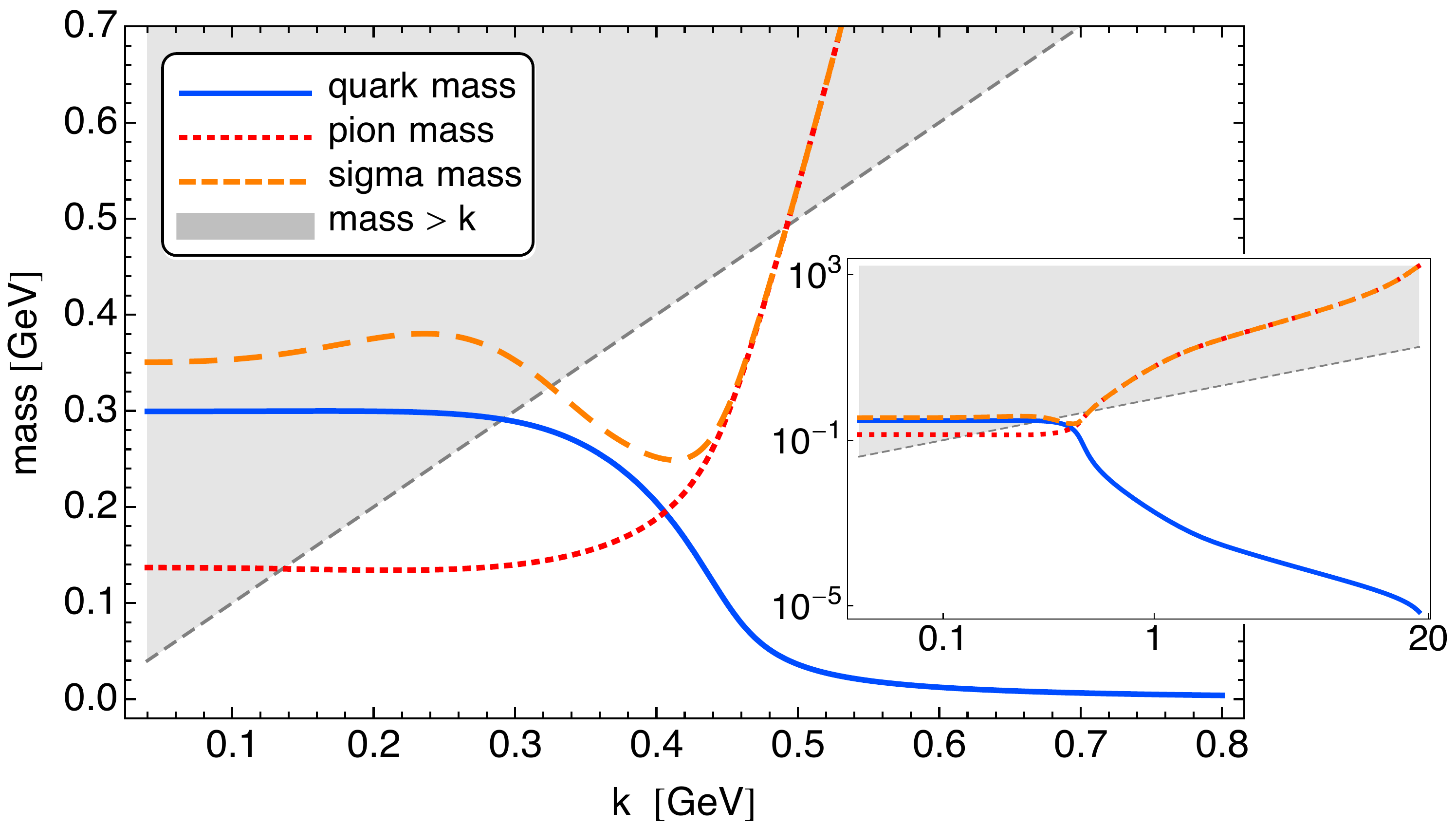}
  \caption{The renormalized quark, pion and sigma masses as a function
  of the RG scale. The inset figure shows the masses for a larger range
  of scales. The shaded gray area indicates which fields contribute dynamically:
  masses within the gray area exceed the cutoff scale and the
  corresponding fields are therefore decoupled from the dynamics. On
  the other hand, fields with masses within the white area are
  dynamical.}\label{fig:masses}
\end{center}
\end{figure}

In \Fig{fig:Gprops} we show the quenched and unquenched gluon
propagators. The quenched gluon propagator is a FRG input from
\cite{Fischer:2008uz,FP}. We clearly see that the screening effects of
dynamical quarks decrease the strength of the gluon propagator.
\Fig{fig:Gprops} also shows the partially unquenched results (denoted
by ``QCD (reduced)" in \Fig{fig:Gprops}) for the propagator. Here,
partially unquenched refers to an approximation, where the gluon
propagator is a direct sum of Yang-Mills propagator and vacuum
polarisation, see Eq.~\eq{eq:eta-simple}. It shows deviations from the
fully unquenched computation. This is seemingly surprising as it is
well-tested that partial unquenching works well even at finite
temperature, see
e.g. \cite{Braun:2009gm,Pawlowski:2010ht,Fischer:2011mz,%
  Fischer:2012vc,Fischer:2013eca,Fischer:2014vxa}. However, we first
notice that the importance of quark flucutations is decreased at
finite temperature due to the Matsubara gapping of the quarks relative
to the gluons.  This improves the reliability of the partial
unquenching results. Moreover, in these works the infrared strength is
phenomenologically adjusted with the constituent quark mass in the
vacuum. This effectively accounts for the difference between
unquenching and partial unquenching. Note that this finding rather
supports the stability and predictive power of functional approaches.

On the other hand this also entails that the full unquenching
potentially is relevant in situations where the vacuum balance between
pure glue fluctuations and quark fluctuations is changed due to an
enhancement of the quark fluctuations. Prominent cases are QCD with a
large number of flavours, and in particular QCD at finite
density. Indeed, \eq{eq:gprop} even shows the self-amplifying effect
at large quark flucutations: The sign of the correction by $\Delta
\eta_{A,k}$ is such that when it grows large, the ratio $\alpha^{\
}_{s,\rm QCD}/\alpha^{\ }_{s,\rm YM}$ decreases as does $\eta_{\rm
  glue}$ and the importance of the matter fluctuations is further
increased. A more detailed study of this dynamics in the above
mentioned situations is deferred to a subsequent publication.

\begin{figure}[t]
\begin{center}
  \includegraphics[width=.95\columnwidth]{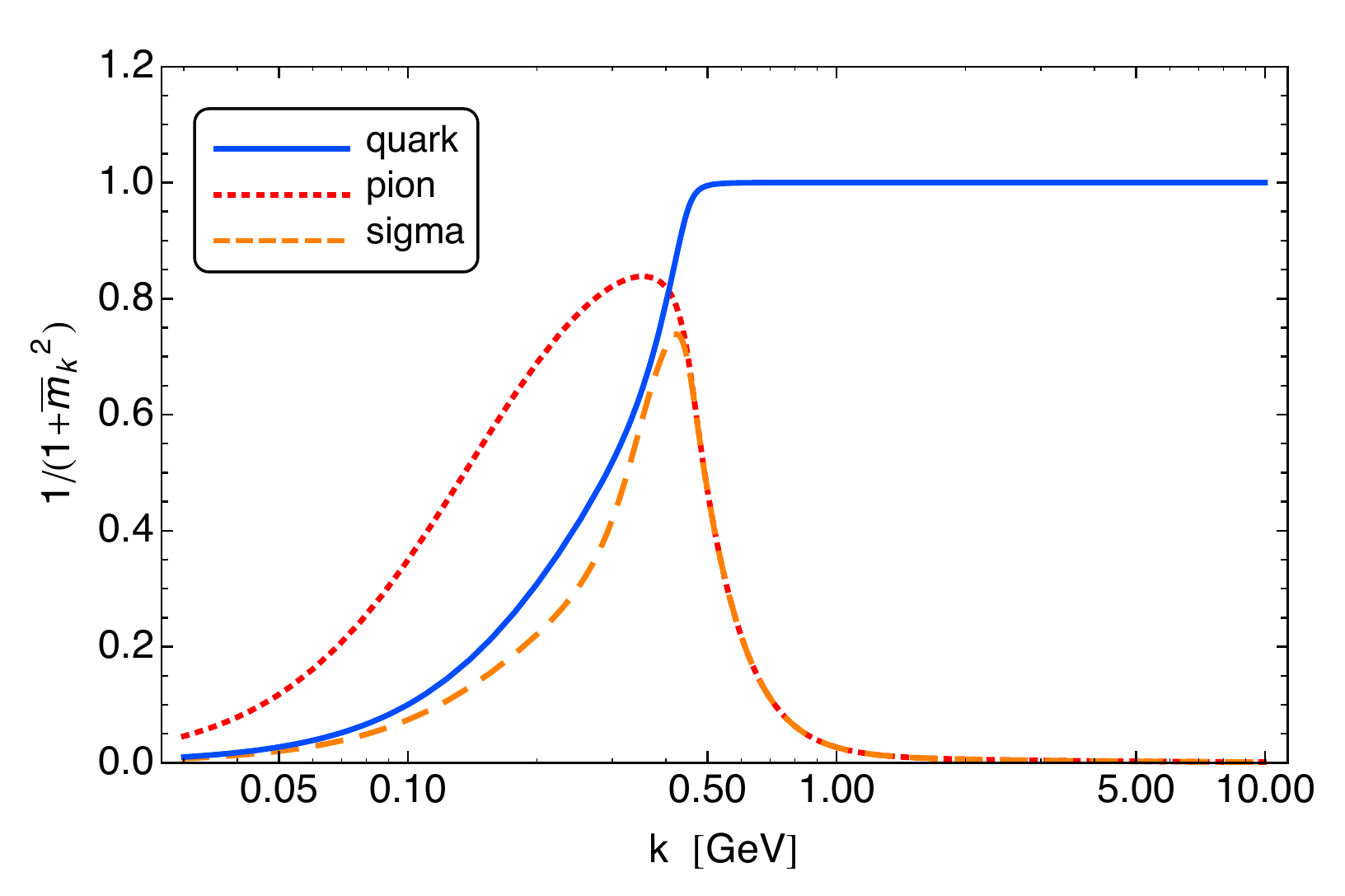}
  \caption{Dimensionless RG-invariant propagators as functions of the RG scale.}\label{fig:props}
\end{center}
\end{figure}

Using the same parameters as in Ref.~\cite{Mitter:2014wpa}, we compare
the quenched and unquenched quark propagators in
Fig.~\ref{fig:qprop}. As for the gluon propagator,
Fig.~\ref{fig:Gprops}, we see large unquenching effects. Unquenching
results in smaller quark masses (blue lines) and larger wave function
renormalizations $Z_{q,k}$, and, therefore, enhanced quark
fluctuations, as expected. Furthermore, we see that the generation of
constituent quark masses takes place at smaller scales in the
unquenched case. This can again be traced back to screening effects:
The effects of gauge fluctuations are suppressed in the presence of
dynamical quarks and lead to weaker gauge couplings. Since the
strength of the gauge couplings triggers chiral symmetry breaking,
criticality of the four-quark interactions is reached later in the
flow for weaker gauge couplings. Hence, chiral symmetry breaking takes
place at smaller scales in the presence of dynamical quarks.

The results for the different running gauge couplings
  $\alpha_{\bar q A q}$, $\alpha_{\bar c A c}$ and $\alpha_{A^3}$
  discussed in Sec.~\ref{sec:gauge} are shown in \Fig{fig:alphas}. At
  scales $k\gtrsim 3~\text{GeV}$ they agree with the perturbative
  running. This constitutes a non-trivial consistency of the present
  computation. At lower scales, non-perturbative effects induce
  different runnings.

\begin{figure}[t]
\begin{center}
  \includegraphics[width=.92\columnwidth]{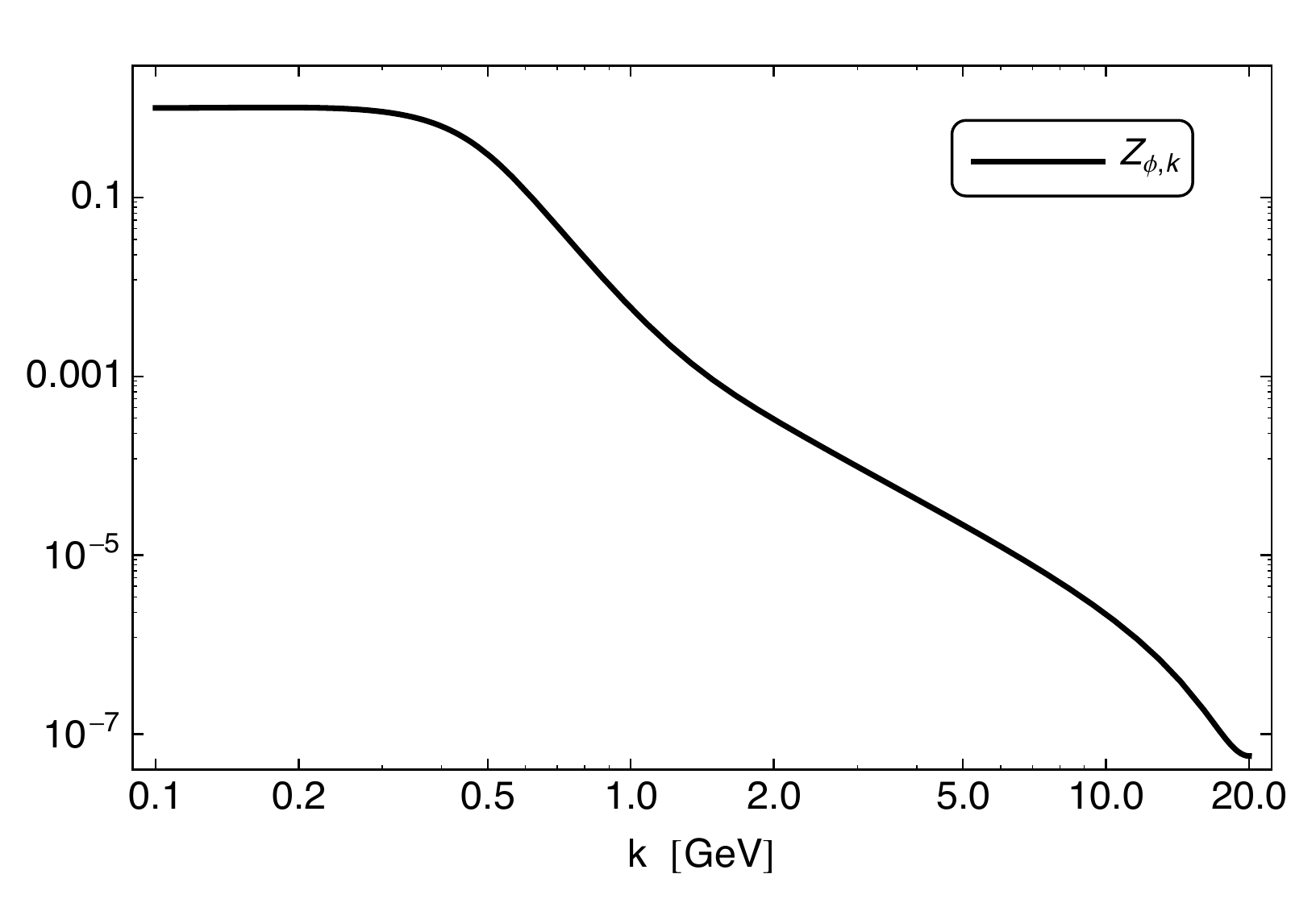}
  \caption{Wave-function renormalization of the mesons.}\label{fig:zphi}
\end{center}
\end{figure}

  The different strengths of the gauge couplings in the
  non-perturbative regime is a direct consequence of the mass gap that
  develops in the gluon dressing function $Z_{A,k}$. Owing to our
  construction for the vertices and the gluon propagator,
  \eq{eq:vertices} and \eq{eq:simpleA}, all non-trivial informations
  about the gauge sector are encoded in the gauge couplings. In
  particular, they genuinely involve powers of $Z_{A,k}^{1/2}$ that
  correspond the number of external gluon legs attached to
  them. Hence, the more external gluonic legs the coupling has, the
  more its strength is suppressed by the emerging gluon mass gap. This
  explains why the three-gluon vertex $\alpha_{A^3}$ is much weaker in
  the non-perturbative regime than $\alpha_{\bar q A q}$ and
  $\alpha_{\bar c A c}$: it is suppressed by $Z_{A,k}^{3/2}$, while
  the quark-gluon and ghost gluon couplings are only suppressed by
  $Z_{A,k}^{1/2}$. The gluon dressing function as we defined it here
  diverges for $k\!\rightarrow\! 0$, and, thus, all gauge couplings
  become zero in this limit.

  The fact that $\alpha_{\bar c A c}$ is weaker than $\alpha_{\bar q A
    q}$ can be attributed to the neglected momentum dependencies in
  this sector. Since all diagrams that drive the flow of the
  ghost-gluon vertex are proportional to the external momentum, they
  vanish for our approximation and $\alpha_{\bar c A c}$ only runs
  canonically, see \eq{eq:gcac}. If these momentum-dependencies were
  taken into account, the ghost-gluon vertex would even be stronger
  than the quark-gluon vertex, at least in the quenched case
  \cite{Mitter:2014wpa}.

The present approach allows an easy access to the relative importance
of quantum fluctuations of the respective fields: we find that for the
renormalised, dimensionless mass being larger than one,
\begin{align}\label{eq:imflucs}
\bar m_\Phi^2 = \0{m_\Phi^2}{Z_\Phi\,k^2}\geq  1 \,,
\end{align} 
all threshold functions that depend on the propagator of the
respective field mode are suppressed with powers of $1/\bar
m_\Phi^2$. This entails that the dynamics of the system is not
sensitive to fluctuations of this field.  In turn, for $\bar
m_\Phi^2\leq 1$ the field mode is dynamical. Note that, of course, $\bar
m^2_\Phi=1$ is not a strict boundary for the relevance of the
dynamics. In Figs.~\ref{fig:masses} and \ref{fig:props} we show $\bar
m^2_\Phi$ for the matter fields. In the shaded area the condition
\eq{eq:imflucs} applies, and the respective matter fields do not
contribute to the dynamics. This already leads to the important
observation that the resonant mesonic fluctuations are only important
for the dynamics in a small momentum regime with momenta $p^2 \lesssim
800$ MeV, see also \Fig{fig:props}. While the $\sigma$- and
quark-modes decouple rather quickly at about 300 - 400 MeV, the $\vec
\pi$ as a pseudo-Goldstone mode decouples at its mass scale of about
140 MeV.

\begin{figure}[t]
\begin{center}
  \includegraphics[width=1.\columnwidth]{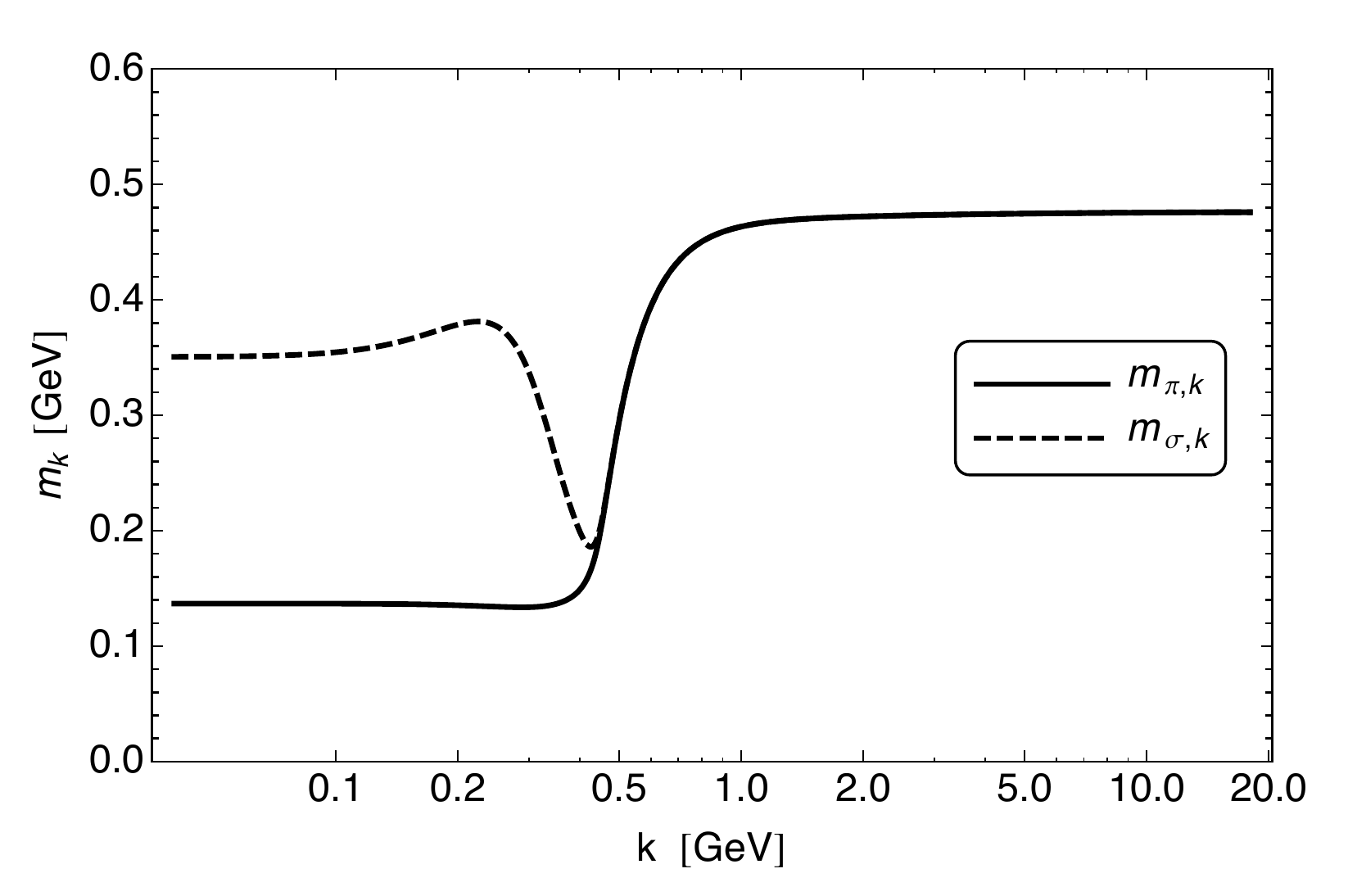}
  \caption{The masses $m_{\pi/\sigma,k} =
    \sqrt{\Gamma^{(2)}_{\sigma/\pi}(0)} = k Z^{1/2}_{\phi,k}
    \bar{m}_{\pi/\sigma,k}$ of the mesons.}\label{fig:baremass}
\end{center}
\end{figure}

In turn, in the ultraviolet regime, the mesonic modes decouple very rapidly,
see \Fig{fig:props} for the size of the propagator measured in units of
the cutoff. At about 800 MeV this ratio is already 0.1 and above this
scale the mesonic modes are not important, and QCD quickly is
well-described by quark-gluon dynamics without resonant interactions.
This observation is complementary to the fact that the initial
condition of the Yukawa coupling does not play a role for the
physics at vanishing coupling, see \Fig{fig:yukawa}. For all initial
cutoff scales $\Lambda\gtrsim 5$ GeV, its initial value is washed out
rapidly, leading to a universal infrared regime with the prediction of
$\bar h$ at $k=0$.

We add that the Yukawa coupling relates to the ratio
between constituent quark mass and the vacuum expectation value of the 
field~$\bar\sigma$,
\begin{align}
\bar h = \0{ \bar m_q}{\bar{\sigma}_0}\,.  
\end{align} 
Note that it cannot be tuned and is a predicition of the theory. On
the other hand, in low-energy model studies, the (renormalised)
quantities $\bar{m}_q$ and~$\bar{\sigma}_0$ corresponding to physical
observables are related to model parameters, and have to be tuned such
that $\bar{m}_q$ and $\bar{\sigma}_0$ assume their physical values.

The decoupling of meson degrees of freedom is also reflected in the
behaviour of the meson wave-function renormalisation $Z_{\phi,k}$
shown in \Fig{fig:zphi}. Starting at scales $k > 500\, \text{MeV}$,
$Z_{\phi,k}$ decreases very rapidly towards the UV. There, it is about
seven orders of magnitude smaller than in the hadronic regime, where
it is $\mathcal{O}(1)$. Furthermore, the masses $m^2_{\pi/\sigma,k} =
\Gamma^{(2)}_{\sigma/\pi}(p^2=0) = Z_{\phi,k} M_{\pi/\sigma,k}^2$
become scale-independent for $k > 800\, \text{MeV}$ as shown in
\Fig{fig:baremass}. This implies that the meson sector becomes trivial
beyond this scale. We see that the drastic decrease of the meson
wave-function renormalisation triggers the large renormalised meson
masses $M^2_{\pi/\sigma,k} = m^2_{\pi/\sigma,k}/Z_{\phi,k}$ shown in
\Fig{fig:masses}, which are responsible for the suppression of the
dynamics of the meson sector at scales $k > 800\, \text{MeV}$. In
turn, this implies that if we start with decoupled mesons in the UV as
in the present case, i.e. initial meson masses much larger than the
cutoff, the running of $Z_{\phi,k}$ drives the meson masses to their
small values in the IR. Without this peculiar behaviour of the meson
wave function renormalisation, the meson masses would never become
smaller than the cutoff scale and hence meson dynamics could not be
generated dynamically. The fact that our results are independent of
the exact value of the initial renormalised meson mass
$M_{\phi,\Lambda} \gtrsim\Lambda$ implies that the running of
$Z_{\phi,k}$ depends on the initial value $M_{\phi,\Lambda}$. Indeed,
if we choose an initial meson mass that is one order of magnitude
smaller (larger), $Z_{\phi,k}$ falls off two orders of magnitude less
(more). This is a direct consequence of the definition of the
renormalised mass, c.f. \eq{eq:mbars} with $M_k = k \bar m_k$, and the
observation that the running of the meson masses is exclusively driven
by $Z_{\phi,k}$ in the UV, cf. \Fig{fig:baremass}. Note that this
behaviour of $Z_{\phi,k}$ has consequences also for low energy models
in the local potential approximation, since for scales larger than
about $800 \, \text{MeV}$, the effect of running wave-function
renormalisations can not be neglected.

Finally, we discuss further consequences of our findings for low energy
effective models.  To that end we note that the gluon modes decouple
at momenta below $500 - 700$ MeV. This is seen from the plot of the
gluon dressing functions, \Fig{fig:Gprops}, as well as that of the
gluonic couplings in \Fig{fig:alphas}. This overlaps with the scale
regime where the mesonic degrees of freedom start to dominate the
dynamics. 

Consequently, low energy effective models aiming at quantitative
precision that do not take into account any glue fluctuations should
be initiated at a UV-scale of about 500 MeV.  In this
regime, however, the quark-meson sector of QCD carries already some
fluctuation information in non-trivial mesonic and quark-meson
couplings. In other words, the standard initial effective Lagrangian
of these models has to be amended by additional couplings. These
couplings, however, can be computed from QCD flows. 

It has been shown in \cite{Helmboldt:2014iya} that in these low energy
effective models thermal fluctuations affect the physics at 
surprisingly large scales, for thermodynamical consequences, see
Ref.~\cite{Herbst:2013ufa}. This is even more so for density fluctuations
that lack the exponential suppression present for thermal
fluctuations. Thus, we conclude that the low UV cutoff scale for
quantitatively reliable low energy effective models enforces the
computation of temperature- and density-dependent initial
conditions. Indeed the same argument holds true for other external
parameters such as the magnetic field.

\section{Conclusions \& Outlook}\label{sec:conc}

In the present work, we have set up a non-perturbative FRG approach to
QCD, concentrating on the effects of a full unquenching of the glue
sector. We also provided a detailed study of the fluctuation physics
in the transition region from the quark-gluon regime to the hadronic
regime. This includes a discussion of the relative importance of the
fluctuations of quark, meson and glue fluctuations. A detailed
discussion is found in the previous section. 

Here we simply summarise the main results. Firstly, we have shown that
the full back-coupling of the matter fluctuations in the glue sector
also plays a quantitative role in the vacuum. In the present
two-flavour case, it accounts for about 10-15\% of fluctuation strength
in the strongly correlated regime at about 1 GeV. This hints strongly
at the importance of these effects in particular at finite density,
where the importance of quark fluctuations is further increased and
the effect is amplified.

Secondly, the still qualitative nature of the present approximation
necessitates the adjustment of the infrared coupling strength, fixed
with the constituent quark mass. However, the inclusion of dynamical
hadronisation which re-enforces the four-fermion running, this
phenomenological tuning is much reduced. In future work we plan to
utilise the findings of the quantitative study \cite{Mitter:2014wpa} in quenched
QCD for improving our current approximation towards quantitative
precision, while still keeping its relative simplicity. 

Finally, we have also discussed how low energy effective models emerge
dynamically within the present set-up due to the decoupling of the
glue sector: the present results and their extensions can be used to
systematically improve the reliability of low energy effective models
by simply computing the effective Lagrangian of these models at their
physical UV cutoff scale of about 500 - 700 MeV. Moreover, the 
temperature- and density-dependence of the model
parameters at this UV scale can be computed within the present set up.  

Future work aims at a fully quantitative unquenched study by also utilising the results 
of \cite{Mitter:2014wpa}, as well as studying the dynamics at finite temperature and density.

{\it Acknowledgments} --- We are greatful to Lisa M. Haas for many 
discussions and collaboration in an early stage of the project. We
thank Tina Herbst, Mario Mitter and Nils Strodthoff for discussions
and collaboration on related projects.  J.B. acknowledges support by
HIC for FAIR within the LOEWE program of the State of Hesse. Moreover,
this work is supported by the Helmholtz Alliance HA216/EMMI and by
ERC-AdG-290623. L.F. is supported by the European Research Council
under the Advanced Investigator Grant ERC-AD-267258.

\appendix

\section{Dynamical hadronisation and low energy effective models}\label{app:lowhad}

In low energy models of QCD, such as (Polyakov-loop enhanced)
Nambu--Jona-Lasinio models or quark-meson models, gluons are considered
to be integrated out and one is left with effective four-quark
interactions, either explicitly or in a bosonised formulation. The
latter is particularly convenient as the phase with spontaneous broken
chiral symmetry is easily accessible. There, the formulation of the
effective theory is usually based on the conventional
Hubbard-Stratonovich bosonization rather than dynamical
hadronisation. Following our arguments given in Sect.~\ref{sec:dynhad},
the question arises whether dynamical hadronisation leads to
quantitative and/or qualitative corrections in the context of low
energy effective model.

Since the matter part of our truncation \eq{eq:trunc} is that of a
quark-meson model, we will consider here the special case of the
quark-meson model defined by switching off all gluon contributions in
\eq{eq:trunc}. To see the effect of dynamical hadronisation, we look
at the ratios of IR observables obtained with and without dynamical
hadronisation. To this end, we choose $\Lambda_\text{LE}=1$~GeV as a typical
UV-cutoff scale and use the same set of initial conditions in both
cases. For results see Tab. \ref{tab:qmhad}.
\begin{table}[h]
\renewcommand{\arraystretch}{1.5}
\begin{tabularx}{1\columnwidth}{ X X X X }
  \hline\hline $f_\pi/\tilde{f}_\pi$ & $M_q/\tilde{M}_q$ &
  $M_\pi/\tilde{M}_\pi$
  & $M_\sigma/\tilde{M}_\sigma$\\
  \hline 0.995 & 0.997 &
  1.003 & $ 0.990 $\\
  \hline\hline
\end{tabularx}
\caption{Effect of dynamical hadronisation on a quark-meson
  model: The quantities with/without a tilde are the results obtain from a solution
  of the flow equations of the quark-meson model with/without
  dynamical hadronisation techniques.}
\label{tab:qmhad}
\end{table}

We see that the effect of dynamical hadronisation on physical
observables of a low-energy quark-meson model (without gluons) is negligible,
since it only gives corrections of less than 1\%. This does
not change if we vary the UV-cutoff within the range of typical values
for this type of models, i.e $\Lambda_\text{LE}\in
[0.5,1.5]$~GeV. Furthermore, it implies in particular that the
mis-counting problem discussed in Sect.~\ref{sec:dynhad} is less severe
in low energy models.

This observation can be understood by looking at the flow of the
four-quark interaction $\lambda_{q,k}$, see Eq.~\eq{eq:dtbarlaq}. In
case of the quark-meson model, only the meson box diagrams $\sim
h_k^4$ contribute to the flow, see also \Fig{fig:box}, while the gluon
box diagrams are neglected. In the chirally symmetric regime, the
mesons are decoupled and the corresponding contributions to the flow
are therefore suppressed. Furthermore, in the hadronic regime, the
quarks acquire a large constituent mass and, in addition, the pions
become light. Therefore, the contribution from dynamical hadronisation
to the flow of the Yukawa coupling \eq{eq:hadh1}, $\sim
\bar{m}_{\pi,k}^2\partial_t\bar{\lambda}_{q,k}$, is suppressed by
these two effects in broken regime. Thus, following our present
results, in particular \Fig{fig:masses}, the only regime where
dynamical hadronisation can play a role in a low-energy model is in
the vicinity of chiral symmetry breaking scale. However, since this
region is small compared to range of scales considered even in
low-energy models, only very small corrections related to the
re-generation of four-quark interactions are accumulated from the RG
flow.

Note, however, that we checked this statement only in vacuum and it
might not be true in medium, especially at large chemical potential
where quark fluctuations are enhanced. This can potentially lead to
larger, non-negligible corrections from dynamical hadronisation.  We
also emphasise that we used the same initial conditions for our
comparison of the RG flow of the quark-meson model with and without
dynamical hadronisation techniques. However, usually the parameters of
low-energy models are fixed in the vacuum, independent of the model
truncation. Once the parameters are fixed, these models are then used
to compute, e.g., the phase diagram of QCD at finite temperature and
chemical potential.  In this case, it may still very well be that the
use of dynamical hadronisation techniques yield significant
corrections.

\section{Flow equations of the couplings}\label{app:flows}

In this appendix, we briefly discuss the derivation of the flow equations of the couplings
before dynamical hadronisation techniques are applied.

We expand the effective potential and the Yukawa coupling about a
fixed expansion point $\kappa$, see \eq{potential}. The advantage of
such an expansion is that it is numerically stable, inexpensive and it
converges rapidly \cite{Pawlowski:2014zaa}. This allows us to take the
full field-dependent effective potential $V_k(\rho)$ and Yukawa
coupling $h_k(\rho)$ into account in the present analysis.

The flow equation of the effective potential including the symmetry
breaking source, $V_k(\rho)-c\sigma$, is obtained by evaluating
\eq{fleq} for constant meson fields, $\phi(x)\rightarrow\phi$ and
vanishing gluon, quark and ghost fields. In this case, the effective
action reduces to $\Gamma_k=\Omega^{-1}(V_k(\rho)-c\sigma)$,
where~$\Omega$ is the space-time volume. The
flow of the effective potential $\bar V_k(\bar\rho)=V_k(\rho)$ is then
given by:
\begin{align}\label{Vflow}
\begin{split}
&\left.\partial_t\right|_\rho\! \bar V(\bar\rho) =\\
&\quad 2 k^4 v(d) \Bigl\{ \left[ (N_f^2-1)l^B_0(\bar{m}_{\pi,k}^2;\eta_{\phi,k})\right.\Bigr.\\
&\quad+\Bigl.\left. l^B_0(\bar{m}_{\sigma,k}^2;\eta_{\phi,k})    \right]-4N_f N_c l^F_1(\bar{m}_{q,k}^2;\eta_{q,k})  \Bigr\}\,,
\end{split}
\end{align}
where $v(d)=(2^{d+1} \pi^{d/2} \Gamma(d/2))^{-1}$ and the treshold functions $l^B_1$ and $l^F_1$ are given in \Eq{eq:lalb}. The flows of the
couplings in (\ref{potential}) can be derived from the above equation
via:
\begin{align}\label{Vnflows}
\begin{split}
&\Bigl.\partial_{\bar\rho}^n\left.\partial_t\right|_\rho\! \bar V(\bar\rho)\Bigr|_{\bar\rho=\bar\kappa_k} =\\
&\quad (\partial_t-n\eta_{\phi,k})\bar\lambda_{n,k}-\bar\lambda_{n+1,k} (\partial_t+\eta_{\phi,k})\bar\kappa_k\,.
\end{split}
\end{align}
Rescaling the expansion point and the symmetry breaking source in
order to formulate RG invariant flows introduces a canonical running
for these parameters:
\begin{align}\label{eq:kandc}
\begin{split}
\partial_t\bar\kappa_k &= -\eta_\phi \bar\kappa_k\,,\\
\partial_t\bar c &= \frac{1}{2}\eta_\phi \bar c\,.
\end{split}
\end{align}
The renormalised minimum of the effective potential
$\bar\rho_{0,k}=\bar\sigma_{0,k}^2/2$, which determines the pion decay
constant at vanishing IR-cutoff, $\bar\sigma_{0,k=0}=f_\pi$, and
serves as an order parameter for the chiral phase transition, is
obtained from:
\begin{align}
\Bigl.\partial_{\bar\rho}\left[ \bar V_k(\bar\rho)-\bar c_k\bar\sigma \right]\Bigr|_{\bar\rho_{0,k}}=0\,.
\end{align}
All physical observables such as $f_\pi$ and the masses are defined at
vanishing cutoff-scale $k=0$ and at the minimum of the effective
potential $\bar\rho=\bar\rho_{0,k=0}$.

We define the field-dependent Yukawa coupling via the relation
$m_{q,k}(\rho)= \sigma h_k(\rho)$ at vanishing external momentum and
constant meson fields, leading to the following projection:
\begin{align}\label{eq: hproj}
\begin{split}
  \partial_t h_k(\rho)=&-\frac{1}{\sigma}\frac{i}{4\Nc\Nf}\lim_{p\rightarrow
      0}\text{Tr}\!\left.\left(
        \frac{\delta^2\partial_t\Gamma_k}{\delta q(-p)\delta\bar{q}(p)}
       \right)\right|_{\rho(x)=\rho}\,.
\end{split}
\end{align}
The resulting flow is given by:
\begin{align}\label{flowYukawa2}
\nonumber&\left.\partial _t\right|_{\bar\rho}\! \bar{h}(\bar\rho) =\\ \nonumber
&\quad\left(\eta_{q,k} +\frac{1}{2}\eta_{\phi,k}\right) \bar{h}_k(\bar\rho)\\ \nonumber
&\quad - v(d) \bar{h}_k(\bar\rho)^3\!
\left[ (N_f^2-1) \, L_{1,1}
^{(FB)}(\bar{M}_{q,k}^2,\bar{m}_{\pi,k}^2;\eta_{q,k},\eta_{\phi,k})\right.\\ \nonumber
&\quad - \left. L_{1,1}^{(FB)}(\bar{m}_{q,k}^2,\bar{m}_{\sigma,k}^2;\eta_{q,k},\eta_{\phi,k})\right]\\ \nonumber
&\quad+8 v(d) \bar h_k(\bar\rho)\,\bar h_k'(\bar\rho)\,\bar\rho\,\bigl[\bar h_k(\bar\rho)+2\bar\rho\bar h_k'(\bar\rho)\bigr]\\ \nonumber
&\quad\times L_{1,1}^{(FB)}(\bar{m}_{q,k}^2,\bar{m}_{\sigma,k}^2;\eta_{q,k},\eta_{\phi,k})\\ \nonumber
&\quad-2v(d)k^2\left[\left(3\bar h_k'(\bar\rho)+2\bar\rho\bar h_k''(\bar\rho)\right)l^B_1(\bar{m}_{\sigma,k}^2;\eta_{\phi,k})\right.\\ \nonumber
&\quad+\left.3\bar h_k'(\bar\rho)l^B_1(\bar{m}_{\pi,k}^2;\eta_{\phi,k})  \right]\\ \nonumber
&\quad -8 (3+\xi) \, C_2(N_c)\,  v(d) \, g_{\bar q A q,k}^2 \, \bar{h}_k(\bar\rho)\\
&\quad\times L_{1,1} ^{(FB)}(\bar{m}_{q,k}^2,0;\eta_{q,k},\eta_{A,k})\,,
\end{align}
$\xi$ is the gauge fixing parameter, which we set to zero since we use
Landau gauge in this work. The function $L_{1,1}^{(FB)}$ is given in \Eq{eq:Ls}. The flows of the renormalised couplings in
(\ref{potential}) are:
\begin{align}\label{eq:hnflow}
\begin{split}
&\Bigl.\partial_{\bar\rho}^n\left.\partial_t\right|_\rho\! \bar h(\bar\rho)\Bigr|_{\bar\rho=\bar\kappa_k} =\\
&\quad (\partial_t-n\eta_{\phi,k})\bar h_{n,k}-\bar h_{n+1,k} (\partial_t+\eta_{\phi,k})\bar\kappa_k\,.
\end{split}
\end{align}
It was shown in Ref.~\cite{Pawlowski:2014zaa}, already a $\phi^4$ expansion
of the effective potential, corresponding to $N_V\!=\!2$ in
\eq{potential} gives quantitatively precise results for small
temperatures and densities. On the other hand, a leading order
expansion of the Yukawa coupling, i.e. $N_h\!=\!0$, is not sufficient
since the expansion is not yet converged. Here, we choose $N_h\!=\!3$
to ensure that we take the effect of the full field-dependent
Yukawa coupling into account. Note that we have to choose $N_V\geq
N_h$ for numerical stability and therefore choose $N_V\!=\!5$.

For the flow of the four-quark coupling we choose the projections in
\cite{Braun:2011pp}. This yields
\begin{align}\label{eq:dtbarlaq}
  \nonumber &\partial_t \bar{\lambda}_{q,k} =\\ \nonumber &- g_{\bar q
    A q,k}^4\, \left(\frac{2N_c^2-3}{N_c}\right) v(d)
  L_{1,2}^{(FB)}(\bar{m}^2_{q,k};\eta_{q,k},\eta_{A,k})\\ \nonumber
  & + \bar{h}_k(\bar\kappa_k)^4\left(\frac{2}{N_c}+1\right) v(d)\\
  &\times
  L_{1,1,1}^{(FB)}(\bar{m}^2_{q,k},\bar{m}^2_{\pi,k},\bar{m}^2_{\sigma,k};\eta_{q,k},\eta_{\phi,k})\,
  .
\end{align}
The treshold functions $L_{1,2}^{(FB)}$ and $L_{1,1,1}^{(FB)}$ are shown in \Eq{eq:Ls}. In Eq.~\eq{eq:dtbarlaq}, we anticipate full dynamical hadronisation
for the four fermi interaction. This leads to a vanishing four-quark
coupling $\bar \lambda_{q,k}=0$ on the right-hand side: the
self-coupling diagram proportional to $\bar \lambda_{q,k}^2 $ is
dropped. Furthermore, we neglect contributions from higher order
quark-meson vertices related to field-derivatives of $\bar
h_k(\bar\rho)$, since they are subleading.

The anomalous dimensions are related to the flow of the wave-function
renormalisations, $\eta=-\partial_t Z / Z$. The $Z$'s on the other
hand encode the non-trivial momentum dependence of the
propagators. Here, as already discussed above, we approximate the full
momentum, scale and field dependence of the anomalous dimensions by
only scale-dependent ones in the leading order expansion in the fields
in analogy to \eq{potential}:
\begin{align}\label{eq:zapprox}
 Z_{\phi,k}(p^2,\rho)=Z_{\phi,k}(\kappa)\quad \text{and}\quad Z_{q,k}(p^2,\rho)=Z_{q,k}(\kappa)\,.
\end{align}
For the meson anomalous dimension, we therefore use the following projection:
\begin{align}\label{etaphiproj}
  \eta_{\phi,k}=-\frac{1}{2 Z_{\phi,k}}\lim_{p\rightarrow
    0}\frac{\partial^2}{\partial |p|^2}
  \text{Tr}\left.\left(\frac{\delta^2\partial_t \Gamma_k}{
        \delta\pi_i(-p)\delta\pi_i(p)}
    \right)\right|_{\rho=\kappa},
\end{align}
where the choice of $i=1,2,3$ does not matter, owing to the $O(3)$ symmetry of the pions. This yields 
\begin{align}\label{etasigma}
\begin{split}
&\eta_{\phi,k}=\\
&\quad8\, v(d) k^{-2}\bar\kappa_k\, \bar U_k''(\bar\kappa_k)^2  
\, {\mathcal M}_{2,2}(\bar{m}_{\pi,k}^2,\bar{m}_{\sigma,k}^2)\\
&\quad+ 2 N_c \, N_f \, v(d) \, \bar{h}_k(\bar\kappa_k)^2 \left[ {\mathcal
    M}_{4}(\bar{m}_{q,k}^2;\eta_{q,k})\right.\\
&\quad\left. + \frac{1}{2}k^{-2}\bar\kappa_k \bar{h}_k(\bar\kappa_k)^2 {\mathcal
    M}_{2}(\bar{m}_{q,k}^2;\eta_{q,k}) \right].
\end{split}
\end{align}
The functions ${\mathcal M}_{2,2}$ and ${\mathcal
    M}_{2/4}$ are defined in \Eq{eq:etatresh}. Note that it is crucial that the functional derivatives in
\eq{etaphiproj} are with respect to the pions, since sigma-derivatives
would contaminate the flow with contributions proportional to $\sigma
Z_{\phi,k}'(\rho)$.

For the anomalous dimension of quarks, we use the projection
\begin{eqnarray}\label{eq:ano}
\eta_{q,k}&=& -\frac{1}{8 \Nf \Nc Z_{q,k}}\\ & &\hspace{-.5cm} \!\times
\lim_{p\rightarrow 0}\frac{\partial^2}{\partial |p|^2}\text{Tr}\left.\left(\gamma_\mu p_\mu \,\frac{\delta^2
    \partial_t \Gamma_k}{\delta q(-p)\delta\bar q(p)} \right)
\right|_{\rho=\kappa}\, ,
\nonumber 
\end{eqnarray} 
which yields 
\begin{align}\label{etaq}
  \nonumber&\eta_{q}=\\ \nonumber &\quad 2\, v(d) \, C_2(N_c) g_{\bar
    q A q}^2 \,\bigl[ (3-\xi) {\mathcal
    M}_{1,2}(\bar{m}_{q,k}^2,0;\eta_{A,k})\bigr.\\ \nonumber
  &\quad-\bigl. 3(1-\xi)\tilde{\mathcal{M}}_{1,1}(\bar{m}_{q,k}^2,0;\eta_{q,k},\eta_{A,k})\bigr]\\
  \nonumber &\quad+\frac{1}{2} \, v(d) \bigl.[ \left( \bar
      h_k(\bar\kappa_k)+ 2 \bar\kappa_k\bar h_k'(\bar\kappa_k)
    \right)^2\bigr. \\\nonumber 
&\quad \bigl. \times \mathcal
   { M}_{1,2}(\bar{m}_{q,k}^2,\bar{m}_{\sigma,k}^2;\eta_{\phi,k})\bigr.\\
  &\quad+ \bigl.(N_f^2-1)\bar h_k(\bar\kappa_k)^2 \,{\mathcal
    M}_{1,2}(\bar{m}_{q,k}^2,\bar{m}_{\pi,k}^2;\eta_{\phi,k})\bigr]\,.
\end{align}
The corresponding threshold functions can be in \Eq{eq:etatresh}.\\[2ex]

Some of the flow equations in this work were derived with the aid of an extension of DoFun \cite{Huber:2011qr} which utilizes Form \cite{2000math.ph..10025V} and FormLink \cite{2012arXiv1212.3522F}. It was developed and first used by the authors of \cite{Mitter:2014wpa}.

\section{Threshold functions}\label{app:thres}
Here, we collect the threshold functions which enter the flow
equations and encode the regulator and momentum dependence of the
flows. Note that it is here, where the substitution $\eta_{\phi,k} \to
\eta_{\phi,k}-2 \dot{\bar B}_k$ has to be made according to
\eq{eq:etashift}.

Throughout this work, we use $4d$ regulator functions of the form:
\begin{align}\label{eq:regdefs}
\begin{split}
R_k^\phi(p^2)&=Z_{\phi,k}\,p^2r_B(p^2/k^2)\,,\\
R_k^q(p^2)&=Z_{q,k}\,\gamma_\mu p_\mu r_F(p^2/k^2)\,,\\
R_k^{A\,,\mn}(p^2)&=Z_{A,k}\,p^2r_B(p^2/k^2) \, \Pi^\bot _{\mu\nu}(p)\,,
\end{split}
\end{align}
with the transverse projector
\begin{equation}
\label{eq:Pi}
\Pi^\bot _{\mu\nu}(p)=\delta_{\mu\nu} -\0{p_\mu p_\nu}{p^2}\,.
\end{equation}
Note that in the approximation at hand the ghost regulator does not enter.
The optimised regulator shape functions $r_{B/F}(x)$ are given by
\cite{Litim:2000ci}:
\begin{align}\label{litR}
\begin{split}
r_B(x)&=\left(\frac{1}{x}-1\right)\Theta(1-x)\,,\\
r_F(x)&=\left(\frac{1}{\sqrt{x}}-1\right)\Theta(1-x)\,.
\end{split}
\end{align}
The threshold functions for the effective potential are 
\begin{align}\label{eq:lalb}
\begin{split}
 l_n^B(\bar{m}_B^2;\eta_B)&=\frac{2 (\delta_{n,0}+n)}{d} \left(
  1-\frac{\eta_B}{d+2}\right) (1+\bar{m}_B^2)^{-(n+1)}\,,\\
l_n^F (\bar{m}_F^2;\eta_F)&=\frac{2 (\delta_{n,0}+n)}{d} \left(
  1-\frac{\eta_F}{d+1}\right) (1+\bar{m}_F^2)^{-(n+1)}\,,
\end{split}
\end{align}
and that for the Yukawa coupling and the four-quark coupling are 
\vfill
\begin{widetext}
\begin{align}\label{eq:Ls}
 \nonumber L_{1,1}^{(FB)}(\bar{m}^2_F,\bar{m}^2_B;\eta_F,\eta_B)&=\frac{2}{d}
  (1+\bar{m}^2_F)^{-1} (1+\bar{m}^2_B)^{-1} \Biggl
  \{\left(1-\frac{\eta_F}{d+1}\right)
  (1+\bar{m}^2_F)^{-1}+\left(1-\frac{\eta_B}{d+2}\right)
  (1+\bar{m}^2_B)^{-1} \Biggr\}\,,\\ \nonumber
  L_{1,2}^{(FB)}(\bar{m}^2_F;\eta_F,\eta_B)&=\frac{2}{d}
  (1+\bar{m}^2_F)^{-2} \Biggl\{2\left(1-\frac{2
      \eta_B}{d+2}\right)-\left(1-\frac{\eta_F}{d+1}\right)+ 2
  (1+\bar{m}^2_F)^{-1}\left(1-\frac{\eta_F}{d+1}\right)\!
  \Biggr\}\,,\\
  L_{1,1,1}^{(FB)}(\bar{m}^2_F,\bar{m}^2_{B1},\bar{m}^2_{B2};\eta_F,\eta_B)&=\frac{2}{d}(1+\bar{m}^2_F)^{-2}
  (1+\bar{m}^2_{B1})^{-1} (1+\bar{m}^2_{B2})^{-1}\Biggl\{ \left[
    (1+\bar{m}^2_{B1})^{-1}+ (1+\bar{m}^2_{B2})^{-1}\right]\Biggl.\\ \nonumber
  &\quad\times\Biggl.\left(1-\frac{\eta_B}{d+2}\right)+\left[
    2(1+\bar{m}_F^2)^{-1}-1\right]
  \left(1-\frac{\eta_F}{d+1}\right)\Biggr\}.
\end{align}
\end{widetext}
For the anomalous dimensions, we have
\begin{widetext}
\begin{align}\label{eq:etatresh}
 \nonumber \mathcal{M}_{2}(\bar{m}^2_F;\eta_F) &=\left( 1+\bar{m}^2_F\right)^{-4}\,,\\ \nonumber
  \mathcal{M}_{2,2}(\bar{m}^2_{B1},\bar{m}^2_{B2};\eta_B)&=(1+\bar{m}^2_{B1})^{-2}
  (1+\bar{m}^2_{B2})^{-2}\,\\
  \mathcal{M}_{1,2}(\bar{m}^2_F,\bar{m}^2_B;\eta_F,\eta_B) &=
  \left(1-\frac{\eta_B}{d+1}
  \right) (1+\bar{m}^2_F)^{-1} (1+\bar{m}^2_B)^{-2}\,\\ \nonumber
  \mathcal{M}_{4}(\bar{m}^2_F;\eta_F) &=\left(
    1+\bar{m}^2_F\right)^{-4}+\frac{1-\eta_F}{d-2}\left(
    1+\bar{m}^2_F\right)^{-3}-\left( \frac{1}{4}+\frac{1-\eta_F}{2
      d-4}\right) \left(1+\bar{m}^2_F\right)^{-2}\,\\ \nonumber
  \tilde{\mathcal{M}}_{1,1}(\bar{m}^2_F,\eta_F,\eta_B) &=
  \frac{2}{d-1}\left(1+\bar{m}^2_F\right)^{-1}
  \Biggl\{\frac{1}{2}\left(\frac{2\eta_F}{d}-1\right)+
  \left(1-\frac{\eta_B}{d+1}\right)+\left(1-\frac{2\eta_F}{d}\right)\left(1+\bar{m}^2_F\right)^{-1}\Biggr\}\,.
\end{align}
\end{widetext}
Finally, for the flow of $z_{\bar q A q}$ we use
\begin{widetext}
\begin{align}
\begin{split}
  \CN_{2,1}^{(m)}(\bar{m}^2_F,\bar{m}^2_B;\eta_F,\eta_B) &=
  \frac{1}{d} \left(1-\frac{\eta_F}{d+1}\right)(1+\bar{m}_B^2)^{-1}
  \Bigl\{ 2 \bar{m}_F^2(1+\bar{m}_F^2)^{-3}+(1+\bar{m}_F^2)^{-2} \Bigr\}\\
  &\quad+\frac{1}{d} \left(1-\frac{\eta_B}{d+2}\right)(1+
  \bar{m}_B^2)^{-2} \Bigl\{ \bar{m}_F^2(1+\bar{m}_F^2)^{-2}
  +(1+\bar{m}_F^2)^{-1} \Bigr\}\,,\\
  \CN_{2,1}^{(g)}(\bar{m}^2_F;\eta_F,\eta_A) &=
  \frac{1}{d} \left(1-\frac{\eta_F}{d+1}\right)\bar{m}_F^2(1+\bar{m}_F^2)^{-3}+\frac{1}{2 d} \left(1-\frac{\eta_A}{d+2}
  \right)\bar{m}_F^2(1+\bar{m}_F^2)^{-2}\,,\\
  \CN_{1,2}^{(g)}(\bar{m}^2_F;\eta_F,\eta_A) &= \frac{1}{d+1}
  \left(1-\frac{\eta_F}{d+2}\right)\Bigl\{2\bar{m}_F^2(1+\bar{m}_F^2)^{-2}-(1+\bar{m}_F^2)^{-1}\Bigr\}\\
  &\quad+\frac{4}{d+1}
  \left(1-\frac{\eta_A}{d+3}\right)
  (1+\bar{m}_F^2)^{-1}.
\end{split}
\end{align}
\end{widetext}

\section{Infrared parameter}\label{app:IRstrength}

In our study, we introduced
an ``infrared-strength" function $\varsigma_{a,b}(k)$ which we define as
\begin{align}
\varsigma_{a,b}(k) = 1+ a\, \frac{(k/b)^\delta}{e^{(k/b)^\delta}-1}\,,
\end{align}
with $b>0$ and $\delta>1$. Note that the specific form of $\varsigma_{a,b}(k)$ is
irrelevant for our result as long as it has the properties specified
below. It defines a smooth step function centered around $b$ with
interpolates smoothly between
\begin{align}
\varsigma_{a,b}(k\gg b) = 1\quad\text{and}\quad \varsigma_{a,b}(k\ll b) = 1+a\,.
\end{align}
Thus, for $b=\mathcal{O}(1\,\text{GeV})$, $\varsigma_{a,b}(k)$ gives
an IR-enhancement, while it leaves the perturbative regime
unaffected. We then modify the gauge couplings as
\begin{align}\label{eq:girs}
g_{s,k}\longrightarrow \varsigma_{a,b}(k)\, g_{s,k},
\end{align}
where $g_{s,k} = g_{\bar q A q,k}\,,\, g_{A^3,k}\,,\,g_{\bar c A
  c,k}$. We choose the same parameters $a$ and $b$ for every gauge
coupling. Accordingly, the flow equations of the gauge couplings then are
\begin{align}
\partial_t g_{s,k} \longrightarrow g_{s,k}\,\partial_t \varsigma_{a,b}(k) + \varsigma_{a,b}(k)\, \partial_t g_{s,k}.
\end{align}
We have found that our results do not depend strongly on the precise value of $b$ as long as it is
$\mathcal{O}(1\,\text{GeV})$. To be specific, we choose $b\!=\!1.3\,\text{GeV}$ for $\delta\!=\!3$ in the following.

The parameter $a$ is adjusted such that we get physical constituent quark
masses in the infrared. Here, $a\!=\!0.29$ yields $M_{q,0}\! =\!
299.5\,\text{MeV}$, where $M_{q,k}\! =\! k\bar m_{q,k}$ is the
renormalized quark mass.

Since the results in Ref.~\cite{Mitter:2014wpa} demonstrate that the largest source
for systematic errors of our truncation is rooted in the
approximations that enter the flows of the gauge couplings, a
procedure as discussed above is well-justified.  


\bibliography{qcd-phase2}

\end{document}